\documentclass[superscriptaddress,nofootinbib,twocolumn,prb]{revtex4-1}

\usepackage{amsmath,amssymb}
\usepackage{graphicx}
\usepackage{xcolor}
\usepackage{hyperref}
\usepackage{url}
\usepackage[utf8]{inputenc}
\usepackage{slashed}
\usepackage{subfigure}
\usepackage{slashed,bbm}
\usepackage{graphics,psfrag,epsfig}
\usepackage{dsfont}
\usepackage{setspace}

\def\code#1{\texttt{#1}}

\begin{document}

\title{
Four-loop critical exponents for the Gross-Neveu-Yukawa models}
              
\begin{flushleft}
      \normalsize DESY 17-133
\end{flushleft}
\vskip0.8cm

\author{Nikolai~Zerf}
\affiliation{Institut f\"ur Theoretische Physik, Universit\"at Heidelberg, Philosophenweg 16, 69120 Heidelberg, Germany}

\author{Luminita~N.~Mihaila}
\affiliation{Institut f\"ur Theoretische Physik, Universit\"at Heidelberg, Philosophenweg 16, 69120 Heidelberg, Germany}

\author{Peter Marquard}
\affiliation{Deutsches Elektronen Synchrotron (DESY), Platanenallee 6, Zeuthen, Germany}

\author{Igor~F.~Herbut}
\affiliation{Department of Physics, Simon Fraser University, Burnaby, British Columbia, Canada V5A 1S6}

\author{Michael~M.~Scherer}
\affiliation{Institute for Theoretical Physics, University of Cologne, 50937 Cologne, Germany}

\begin{abstract}
We study the chiral Ising, the chiral XY and the chiral Heisenberg models at four-loop order with the perturbative renormalization group in $4-\epsilon$ dimensions and  compute critical exponents for the Gross-Neveu-Yukawa fixed points to order $\mathcal{O}(\epsilon^4)$.
Further, we provide Pad\'e estimates for the correlation length exponent, the
boson and fermion anomalous dimension as well as the leading correction to scaling exponent in
2+1 dimensions. We also confirm the emergence of supersymmetric field theories
at four loops for the chiral Ising and the chiral XY models with $N=1/4$ and $N=1/2$ 
 fermions, respectively.
Furthermore, applications of our results relevant to various quantum transitions in the context of Dirac and Weyl semimetals are discussed, including interaction-induced transitions in graphene and surface states of topological insulators.
\end{abstract}

\maketitle

\section{Introduction}


Critical phenomena near continuous phase transitions constitute one of the cornerstones of our modern understanding of quantum field theory, condensed matter physics and statistical field theory\cite{herbutbook,ZinnJustin:2002ru}.
Near a continuous phase transition the free energy of a physical system exhibits a scaling form\cite{widom1965}, so that the specific heat or the correlation length show a power-law behavior characterized by universal critical exponents.
For example, in a thermal transition, the correlation length diverges as
\begin{align}
	\xi \sim |t|^{-\nu}(1+C|t|^\omega+...)\,,\notag
\end{align}
where $t=(T-T_c)/T_c$ is the reduced temperature measuring the distance from the transition at critical temperature $T_c$. In a quantum phase transition\cite{sachdev2011}, where $T=0$, the reduced temperature is replaced by another measure for the distance from the transition, e.g., the deviation from a critical coupling.
The correlation length exponent~$\nu$ and the subleading exponent~$\omega$ are universal numbers which are identical for a number of phase transitions as specified by symmetry and dimensionality, defining a universality class.

Prime examples for critical behavior are the three-dimensional (3D) $O(N)$ universality classes that can be experimentally studied by various important phase transitions such as the liquid-gas transition in simple gases, the superfluid transition in liquid Helium or the Heisenberg transition in ferromagnets.
From the theoretical side, the development and comparison of different methods has led to an impressive convergence across different theoretical approaches for the $O(N)$ universality classes.
For example, for the Ising universality class there is a three digit agreement for the correlation length exponent $\nu_\mathrm{Ising}\approx 0.630$ across the available theoretical methods\cite{0305-4470-31-40-006,PhysRevE.65.066127,PhysRevE.68.036125,PhysRevB.82.174433,Kos2016} including the  renormalization group (RG), numerical Monte Carlo (MC) approaches and the conformal bootstrap.

In the last years, Dirac and Weyl semimetals\cite{wehling2014,2013arXiv1306.2272V} have emerged as ubiquitous phases of matter in condensed-matter physics, providing prime systems to explore fundamental properties of particles in unprecedented ways, and beyond the realm of high-energy physics.
In Dirac systems the quasi-relativistic energy dispersion leads to universal properties such as, e.g., a linearly vanishing density of states at the Fermi level and the concomitant thermodynamic properties and various response functions.
Under specific circumstances, for example when interactions or disorder are sufficiently strong, Dirac and Weyl systems are believed to undergo second order quantum phase transitions from their semimetallic phase to different types of order\cite{sorella1992,Khveshchenko2001,herbut2006,honerkamp2008,herbut2009,raghu2008,PhysRevB.33.3263,PhysRevB.33.3257,PhysRevLett.107.196803,maciejko2014,PhysRevLett.112.016402,PhysRevB.94.220201,PhysRevB.93.155113}.
Relevant order parameters cover a broad range of suggestions, for example staggered density wave phases, antiferromagnetic states, superconducting orders and more exotic phases.

The critical behavior of a universality class is governed by the dimensionality, symmetry and relevant degrees of freedom of a physical system.
While the critical behavior of the three-dimensional $O(N)$ universality
classes can be conveniently described in terms of  purely bosonic field theories, the presence of symmetry-compatible chiral fermions, as in Dirac and Weyl systems, severely modifies the critical exponents and therefore defines a novel universality class\cite{Rosenstein:1993zf}.
At present, experimental realizations of these quantum transitions are still lacking.
However, in systems like graphene, artificial graphene or cold atoms, related transitions have already been studied\cite{gomes2012,Gutierrez2016} and it can be expected that these quantum phase transitions will be accessible in the near future.

From a general point of view, these Dirac systems close to a phase transition can effectively be described in terms of quasi-relativistic chiral Dirac fermions coupled to a bosonic order parameter which -- depending on the transition -- can have different numbers of components and symmetries.
This defines a general class of Gross-Neveu-Yukawa (GNY) models.
For example, the simplest of these models -- the 3D {\it chiral Ising model} -- with one real scalar field describes the universality class of the interaction-induced quantum transition toward a charge density wave (CDW) of electrons on the 2D honeycomb lattice that breaks the (Ising) sublattice symmetry\cite{herbut2006}.

A precise determination of the universality classes of the 3D GNY models in terms of quantitative critical exponents has been prevented for quite some time due to the lack of suitable methods.
Recently, however, there have been various developments that encourage to pick up on that task again:
\begin{itemize}
	\item Numerical approaches have found a fermion representation allowing for a sign-problem free calculation of the semimetal-to-CDW and related transitions of fermions on the honeycomb lattice\cite{PhysRevD.88.021701,1367-2630-16-10-103008,1367-2630-17-8-085003,PhysRevB.93.155157,Huffman:2017swn}. 
	\item The conformal bootstrap has been developed to determine critical exponents for the $O(N)$ models to unprecedented precision and is now extended to fermionic systems\cite{Poland:2016chs,Bashkirov:2013vya,Iliesiu:2015akf,Iliesiu:2017nrv}.
	\item Non-perturbative field-theoretical methods like the functional renormalization group (FRG) have  achieved the maturity to provide quantitative estimates for critical exponents\cite{PhysRevB.89.205403,Vacca:2015nta,PhysRevB.94.245102,Gies:2017tod,Knorr:2017yze}.
	\item The perturbative renormalization group (pRG) has been formalized to a level that allows for feasible higher-loop calculations for these models\cite{Gracey:2016mio,Zerf:2016fti,Mihaila:2017ble}.
\end{itemize}
However, despite this recent progress in MC simulations and the application of field-theoretical methods, the discrepancies between the results for the GN critical exponents have not been resolved and differences still show up in the first relevant digits.
Concerning the pRG it can be stated that, with a few exceptions\cite{Gracey:2016mio,Zerf:2016fti,Mihaila:2017ble}, most of the universality classes of the GNY models are only known up to two-loop order and no information about the behavior of higher-loop orders is available. This leaves quite some room for improvement on the estimates for critical exponents coming from the pRG.

In this work we considerably extend on previous calculations of critical exponents within the pRG by providing the full analytical expressions for the beta and gamma functions for three representative Gross-Neveu-Yukawa models for general number of fermion flavors~$N$ at four-loop order in $4-\epsilon$ dimensions.
The models are chosen to represent a class quantum transitions relevant to two-dimensional chiral Dirac systems and will be introduced in the next section.
We calculate the inverse correlation length exponent $\nu^{-1}$, anomalous dimensions $\eta_\phi,\eta_\psi$ and the subleading exponent $\omega$ to order~$\mathcal{O}(\epsilon^4)$ and present numerical estimates for the most relevant cases.

The rest of the paper is organized as follows. After introducing the three different models in Sec.~\ref{sec:models}, we specify the RG procedure and the employed computer algebraical tools in Sec.~\ref{sec:RG}. In Sec.~\ref{sec:beta} we present the full set of four-loop RG functions for each of the models. Critical exponents to order $\epsilon^4$ are presented in Sec.~\ref{sec:results}, where we also discuss applications and numerical evaluations of our results and employ simple Pad\'e resummations for the universal critical exponents. Finally, we draw our conclusions. Lengthy expressions for the four-loop contributions are given in a set of appendices.

\section{Models and Applications}\label{sec:models}

Interacting Dirac fermions in two spatial dimensions can undergo a variety of quantum phase transitions towards ordered states with different symmetry breaking patterns\cite{sorella1992,Khveshchenko2001,herbut2006,honerkamp2008,herbut2009,raghu2008,PhysRevB.33.3263,PhysRevB.33.3257,PhysRevLett.107.196803,maciejko2014,PhysRevLett.112.016402,PhysRevB.94.220201,PhysRevB.93.155113}. 
While there is a huge variety of different possible states accompanied by individual subtleties, the general universal critical behavior can be captured by a general class of relativistic Gross-Neveu-Yukawa models.
In this description, the Dirac fermions couple to the order parameters via Yukawa couplings and the order parameters are written in terms of bosonic fields with a corresponding number of components and symmetries.
More explicitly, we discuss three specific models\cite{Rosenstein:1993zf}:
\begin{enumerate}
	\item The {\it chiral Ising model} where chiral Dirac fermions couple to a single-component real-valued order parameter with a discrete $Z_2$ symmetry.
	\item The {\it chiral XY model} where chiral Dirac fermions undergo continuous $U(1)$ symmetry breaking as described by a complex order parameter. This model is closely related to the bosonized version of the Nambu-Jona-Lasinio (NJL) model\cite{doi:10.1093/ptep/ptw120}.
	\item The {\it chiral Heisenberg model} where $SU(2)$ symmetry is broken. Here, the chiral Dirac fermions couple to an order parameter which is represented by a three-component vector.
\end{enumerate}
Explicitly, we describe the quantum critical points of interacting Dirac semimetals in 2+1 dimensions by the following general form of the total action,
\begin{align}\label{eq:lag}
S=\int d\tau d^{D-1}x \, ( \mathcal{L}_\psi + \mathcal{L}_{\psi\phi}+\mathcal{L}_\phi)\,.
\end{align}
The first term in the action is the fermionic kinetic term in Euclidean spacetime
\begin{align}\label{eq:Dirac}
	\mathcal{L}_\psi=\bar\psi(x)\slashed{\partial} \psi(x)\,,
\end{align}
where we define $\slashed{\partial}=\gamma_\mu\partial_\mu$ and use a four-dimensional representation of the Clifford algebra, i.e. $\{\gamma_\mu,\gamma_\nu\}=2\delta_{\mu\nu}\mathds{1}_4$, with $\mu, \nu, = 0,1,...D-1$.
The results on the RG beta and gamma functions are independent from the
explicit choice of the representation of the Clifford algebra as their
derivation only makes use of the anticommutation relation and the trace over
the identity matrix. Therefore, we do not give an explicit representation, here and in the following, and only note that different physical applications come with various explicit representations of the Clifford algebra, see for example Refs.~\onlinecite{Roy:2013,Zerf:2016fti}.
The conjugate of the Dirac field is given by $\bar\psi=\psi^\dagger\gamma_0$.
We generalize the model by introducing a number of $N$ fermion flavors of the four-component Dirac fermions, i.e. the fermion also carries a flavor index $i$  where $\psi = \psi_i$ and $i \in \{1,...,N\}$.
For notational simplicity, we will suppress the flavor index in the following.

Further, we will have to define the Yukawa interaction represented by $\mathcal{L}_{\psi\phi}$ and the purely bosonic part $\mathcal{L}_{\phi}$ including a boson kinetic term and interactions.
In the following, we employ Lorentz-symmetric kinetic terms in all considered Lagrangians, i.e. we set the boson velocity to the same value as the Fermi velocity $v_B=v_F=1$.
This generally provides us with a Lorentz-invariant form of the total action, which, however, is not dictated a priori as the effective GNY models considered here, typically have their origin in a non-Lorentz-invariant lattice description.
In fact, the Lorentz symmetry near fermionic quantum critical points with equal velocities for fermions and bosons has been argued to emerge naturally in the deep infrared regime in a large class of Yukawa theories of the same kind~\cite{},  even if $v_F\neq v_B$ on intermediate scales\cite{Anber:2011xf,Jian:2014pca,Roy:2015zna}.

\subsection{Chiral Ising model}

The first model of the class of Gross-Neveu-Yukawa theories we discuss is the chiral Ising model.
It is represented by the Lagrangian
\begin{align}\label{eq:lagrangian}
	\mathcal{L}_{\chi\text{I}}=\mathcal{L}_\psi+g\phi\bar\psi\psi+\frac{1}{2}\phi(m^2-\partial_\mu^2)\phi+\lambda\phi^4\,,
\end{align}
and includes a real scalar field with one component $\phi$. The model can be considered to result from a Hubbard-Stratonovich decoupled four-Fermi interaction and lies in the same universality class as the purely fermionic GN model~\cite{PhysRevD.10.3235} for (space-time) dimensions $2 < D < 4$. The Lagrangian in Eq.~\eqref{eq:lagrangian} is renormalizable in $D=4-\epsilon$ dimensions. 
The scalar field couples to the fermions with the Yukawa coupling $g$ and has a quartic  coupling $\lambda$.

This version of the Gross-Neveu-Yukawa models has a number of interesting applications depending on the number of fermion flavors $N$. 
For an eight-component spinor $\psi$ ($N=2$) it describes the quantum critical point of the semimetal-insulator transition in graphene, where the ordered state corresponds to a sublattice symmetry broken insulating state with charge order -- the CDW order\cite{herbut2006}.
In the case $N=1$, we deal with a system that lies in the same universality class as spinless fermions on the honeycomb lattice with strong repulsive interactions, also undergoing a semimetal-insulator transition which has been intensely studied, recently, by a broad range of different methods, i.e. quantum Monte Carlo (QMC) simulations\cite{PhysRevD.88.021701,1367-2630-16-10-103008,1367-2630-17-8-085003,PhysRevD.88.021701,Huffman:2017swn,PhysRevB.93.155157}, the Functional Renormalization Group\cite{Vacca:2015nta,PhysRevB.89.205403} (FRG), perturbative RG approaches\cite{doi:10.1093/ptep/ptw120,Mihaila:2017ble} and the conformal bootstrap\cite{Iliesiu:2017nrv}.
For $N=1/4$, it has been argued that supersymmetry emerges at the quantum critical point which might be relevant at the boundary of a topological phase as discussed in Ref.~\onlinecite{Grover:2013rc}.
Finally, the replica limit of this GNY model, $N\to 0$, was suggested to describe the transition from a relativistic semimetallic state to a diffusive metallic phase in 3D Weyl semimetals\cite{PhysRevB.94.220201}.

\subsection{Chiral XY model}

The second model we discuss is the chiral XY model where the chiral Dirac fermions $\psi$ undergo continuous U(1) symmetry breaking as described by the complex order parameter $\phi=\phi_1+i\phi_2$.
The complete Lagrangian is decomposed as
\begin{align}
\mathcal{L}_{\chi\text{XY}}=\mathcal{L}_{\psi}+\mathcal{L}_{\phi, \chi\text{XY}}+\mathcal{L}_{\psi\phi,  \chi\text{XY}}
\end{align}
with the fermionic part $\mathcal{L}_{\psi}$, cf. Eq.~\eqref{eq:Dirac}. The bosonic part of the action reads
\begin{align}
	\mathcal{L}_{\phi, \chi\text{XY}}=|\partial_\mu \phi|^2+m^2|\phi|^2+\lambda|\phi|^4\,,
\end{align}
and the Yukawa interaction $\mathcal{L}_{\psi\phi}$ is 
\begin{align}
	\mathcal{L}_{\psi\phi, \chi\text{XY}}&=g\,\bar\psi\left(\phi_1+i \gamma_5\phi_2\right)\psi\nonumber\\
	&=g\left(\phi\bar\psi P_+\psi+\phi^\ast\bar\psi P_- \psi\right)\,,
\label{eq:XY}
\end{align}
where $P_\pm=\frac{1}{2}(1\pm\gamma_5)$.

Applications of this model in the condensed-matter context can be found in the quantum critical behavior of superconducting states in graphene where the number of fermion flavors is $N=2$, see, e.g., Ref.~\onlinecite{Roy:2013}, where also an explicit choice for the Clifford algebra is discussed.
Further, the case $N=2$ is relevant to a Kekul\'e valence bond solid transition in graphene\cite{hou2007,ryu2009,roy2010c} which is described by a complex order parameter, however, with a discrete $Z_3$ symmetry. 
In the this scenario it was argued that at the QCP the $Z_3$ gets enhanced to an emergent $U(1)$ symmetry leading to a set of critical exponents that is shared with the $N=2$ chiral XY model\cite{li2015,Scherer:2016zwz,Classen:2017hwp}.
Another intriguing scenario where the chiral XY model is relevant are the surface states of three-dimensional topological insulators where emergent supersymmetry has been conjectured at the quantum critical point\cite{lee2007,Roy:2013,Zerf:2016fti}. This implies a field content with fermion flavor number $N=1/2$.
The chiral XY model shares the symmetries of the bosonized version of the Nambu-Jona-Lasinio (NJL) model, also referred to as the Nambu-Jona-Lasinio-Yukawa (NJLY) model, which has recently been discussed in Ref.~\onlinecite{doi:10.1093/ptep/ptw120}.

\subsection{Chiral Heisenberg model}

One of the best candidates for an interaction-induced semimetal-insulator transition of the electronic quasi-particles in graphene ($N=2$) is the transition towards an antiferromagnetic spin-density wave (AF-SDW) state\cite{sorella1992,herbut2006,honerkamp2008,PhysRevX.3.031010,PhysRevLett.111.056801,Buividovich:2016tgo,Otsuka:2015iba} which has been suggested to be accessible by application of biaxial strain\cite{PhysRevLett.115.186602,sanchez2016b}.
In the low-energy effective field-theoretical description this corresponds to a SU(2) symmetry breaking transition with a Heisenberg order parameter field $\vec{\phi}$ having three real components.
The corresponding full model is referred to as the chiral Heisenberg model\cite{Rosenstein:1993zf,PhysRevB.89.205403} with the bosonic kinetic Lagrangian explicitly reading
\begin{align}
	\mathcal{L}_{\phi, \chi\text{H}}=\frac{1}{2}\vec{\phi}\left(m^2-\partial_\mu^2\right)\vec{\phi}+\lambda\left(\vec{\phi}\cdot\vec{\phi}\right)^2\,.
\end{align}
Accordingly, the Yukawa coupling is written as
\begin{align}
	\mathcal{L}_{\psi\phi, \chi\text{H}}=g\vec{\phi}\,\bar\psi\left(\vec{\sigma}\otimes\mathbbm{1}_{2N}\right)\psi\,.
\end{align}
Similar to the other models, the ordered phase of the chiral Heisenberg model is characterized by a finite expectation value of the bosonic field which here corresponds to the spontaneous breaking of spin-rotational symmetry.
Note, that we have directly introduced the generalization to arbitrary number of fermion flavors in the Yukawa interaction. With this generalization, $\psi$ and $\bar\psi$ have $2N$ components for each spin projection and the graphene case is covered by $N=2$.
The explicit implementation of the flavor number is straightforward, as for the derivation of the results only the Clifford algebra and the product $d_\gamma N$ is required, where $d_\gamma$ is the dimension of the representation of the gamma matrices.

\section{Renormalization group and technicalities}\label{sec:RG}

For the renormalization group analysis in $4-\epsilon$ dimensions, we introduce the bare Lagrangian. To that end, we replace the fields and couplings in the Lagrangian from Eq.~\eqref{eq:lag} with their bare counterparts
\begin{align}
\psi\to \psi_0,\quad \phi\to \phi_0,\quad g\to g_0,\quad \lambda \to \lambda_0\,.
\end{align}
We discuss the explicit construction for the chiral Ising model and note that the constructions for the chiral XY and the chiral Heisenberg model work accordingly. The renormalized chiral Ising model Lagrangian reads
\begin{align}
	\mathcal{L}=&Z_\psi\bar\psi\slashed{\partial}\psi-\frac{1}{2}Z_\phi(\partial_\mu\phi)^2+Z_{\phi^2}\frac{m^2}{2}\phi^2\nonumber\\
	&+Z_{\phi\bar\psi\psi} g \mu^{\epsilon/2}\phi\bar\psi\psi+Z_{\phi^4}\lambda\mu^\epsilon\phi^4\,.
\end{align}
Here, we have introduced the energy scale $\mu$ parametrizing the RG flow. The wave function renormalization constants $Z_\psi$ and $Z_\phi$ relate the bare and the renormalized Lagrangian by rescaling the fields according to $\psi_0=\sqrt{Z_\psi}\psi$ and $\phi_0=\sqrt{Z_\phi}\phi$.
For the integration over $D=4-\epsilon$ dimensional spacetime we introduce the rescaling
\begin{align}
g^2\,\to g^2\mu^\epsilon,\quad \lambda\to\lambda\,\mu^\epsilon\,,
\end{align}
which leads to explicit $\mu$ dependencies in $\mathcal{L}$. 
For notational simplicity, we further introduce the squared Yukawa coupling $y=g^2$, see Ref.~\onlinecite{Mihaila:2017ble}. 

The RG scale dependence of the renormalized quantities can be derived from the following relations between the bare and the renormalized mass term, the Yukawa coupling and the quartic coupling, 
\begin{align}\label{eq:Z}
	m^2&=m_0^2Z_\phi Z_{\phi^2}^{-1}\,,\\
	y&=y_0\mu^{-\epsilon}Z_\psi^2Z_\phi Z_{\phi\bar\psi\psi}^{-2}\,,\quad \lambda=\lambda_0\mu^{-\epsilon}Z_\phi^2Z_{\phi^4}^{-1}\label{eq:Z2}\,.
\end{align}
Employing a chain of sophisticated tools developed for higher-loop calculations in the context of the Standard Model of Particle Physics, we evaluate the renormalization group constants
\begin{align}
 Z_\psi,\quad Z_\phi,\quad Z_{\phi^2},\quad Z_{\phi\bar\psi\psi},\quad Z_{\phi^4}\,,
\end{align}
up to four-loop order.
Therefore, we use dimensional regularization (DREG) and the modified minimal subtraction scheme ($\overline{\text{MS}}$).
The tool chain of computer programs operates as follows:
\begin{enumerate}
\item \code{QGRAF}\cite{NOGUEIRA1993279} generates the complete sets of Feynman diagrams.
\item  \code{q2e} and \code{exp}\cite{Harlander1998125,Seidensticker:1999bb} are used to map all Feynman diagrams on one-scale massive tadpole integral topologies and to generate diagram source files.
\item \code{FORM}\cite{Vermaseren:2000nd,Kuipers20131453,Ruijl:2017dtg} is used to process the diagram source files. 
It performs the traces over the Clifford algebra, reduces the $SU(2)$ color amplitudes (in case of the chiral Heisenberg model) with the package \code{COLOR}\cite{vanRitbergen:1998pn}
and rewrites the amplitudes in terms of massive tadpole integrals with different powers of propagators.
Finally it replaces all integrals by their tabulated reduction to a set of
nineteen known master integrals\cite{Czakon:2004bu}.
\item The reduction to master integrals is performed by \code{Crusher}\cite{crusher} and relies on integration-by-parts identities relating integrals with different propagator powers through a system of coupled equations to each other. 
The system of equations can be solved with the Laporta algorithm \cite{laporta:2000} such that all appearing integrals can be written in terms of a linear combination of a finite number of master integrals.
\end{enumerate}

For the computation of the renormalization constants  we employ the method
introduced in Ref.~\onlinecite{Chetyrkin:1997fm}. Explicitly, we assign a mass
regulator to all propagators  and reduce the calculation to the evaluation of
one scale tadpole topologies. Up to three loops we checked our  results against
\code{MATAD}~\cite{STEINHAUSER2001335}.

The total number of diagrams calculated at four-loop level is 31671 for the
chiral Ising and Heisenberg model. For the chiral XY model we implemented two
independent setups using the Feynman rules that can be derived from the first
or the second line of Eq.~\eqref{eq:XY}, i.e. employing a complex scalar
and a Dirac fermion or a real scalar representation and left and right handed Weyl fermions.
The total number of diagrams amounts to 188531 for the first setup and to 7384
for the second one whereas the results for the renormalization constants
completely agree. This is a nontrivial check for our setups up to four loops.    




%

\section{Beta and gamma functions}\label{sec:beta}

The beta functions for the squared Yukawa coupling $y$ and the quartic scalar coupling $\lambda$ are defined as the logarithmic derivatives with respect to the scale $\mu$ to be
\begin{align}
\beta_y=\frac{d\,y}{d\ln \mu},\quad
\beta_\lambda=\frac{d\lambda}{d\ln \mu}\,.
\end{align}
The relation to the renormalization constants is derived from Eqs.~\eqref{eq:Z}-\eqref{eq:Z2} and we work with rescaled couplings $y/(8\pi^2) \to y$ and $\lambda/(8\pi^2)\to \lambda$.
For the Yukawa coupling and the quartic scalar coupling at four-loop order we expand the full expressions according to the scheme
\begin{align}
	\beta_{y, \text{X}}=&-\epsilon y+\beta_{y, \text{X}}^{\text{(1L)}}+\beta_{y, \text{X}}^{\text{(2L)}}+\beta_{y, \text{X}}^{\text{(3L)}}+\beta_{y, \text{X}}^{\text{(4L)}}\,,\\[3pt]
	\beta_{\lambda, \text{X}}=&-\epsilon  \lambda +\beta_{\lambda, \text{X}}^{\text{(1L)}}+\beta_{\lambda, \text{X}}^{\text{(2L)}}+\beta_{\lambda, \text{X}}^{\text{(3L)}}+\beta_{\lambda, \text{X}}^{\text{(4L)}}\,,
\end{align}
where we have defined the functions $\beta_{x, \text{X}}^{\text{(}i\text{L)}}$ with $x\in \{y,\lambda\}$ specifying the coupling, $i\in\{1,2,3,4\}$ specifying the contribution to the flow of the coupling $x$ at loop order $i$ and X specifying the considered model $\text{X} \in \{\chi\text{I}, \chi\text{XY}, \chi\text{H}\}$.
Further, the anomalous dimensions are defined as the logarithmic derivatives
of the wave function renormalizations of the fermion and the boson fields and
of the quadratic operator $\phi^2$, i.e.
$
\gamma_{x}=d\ln Z_{x}/d\ln \mu
$
for $x\in \{\psi,\phi,\phi^2\}$ and read
\begin{align}
	\gamma_{\psi, \text{X}}&=\gamma_{\psi, \text{X}}^{\text{(1L)}}+\gamma_{\psi, \text{X}}^{\text{(2L)}}+\gamma_{\psi, \text{X}}^{\text{(3L)}}+\gamma_{\psi, \text{X}}^{\text{(4L)}},\\
	\gamma_{\phi, \text{X}}&=\gamma_{\phi, \text{X}}^{\text{(1L)}}+\gamma_{\phi, \text{X}}^{\text{(2L)}}+\gamma_{\phi, \text{X}}^{\text{(3L)}}+\gamma_{\phi, \text{X}}^{\text{(4L)}}\,,\\
	\gamma_{\phi^2, \text{X}}&=\gamma_{\phi^2, \text{X}}^{\text{(1L)}}+\gamma_{\phi^2, \text{X}}^{\text{(2L)}}+\gamma_{\phi^2, \text{X}}^{\text{(3L)}}+\gamma_{\phi^2, \text{X}}^{\text{(4L)}}\,.
\end{align}
The beta functions are used to calculate the renormalization group fixed
points and together with the anomalous dimensions provide estimates for universal critical exponents, i.e. the inverse correlation length exponent $\nu^{-1}$, the subleading exponent $\omega$ and the anomalous dimensions of the fermions and the bosons, $\eta_\psi$ and $\eta_\phi$, respectively.

In the next section, we provide the full analytical expressions up to four-loop order for each of the introduced models. We have chosen a normalization of the couplings that allows to easily compare contributions to the beta functions across models. For example, for all models, the one-loop beta functions for the Yukawa coupling and the quartic boson coupling can be written as
\begin{align}
	\beta_{y, \text{X}}^{\text{(1L)}}&=(3-M+2N)y^2,\\
	\beta_{\lambda, \text{X }}^{\text{(1L)}}&=(36+4M)\lambda ^2+4N y \lambda -N y^2\,,
\end{align}
where $M$ is the number of Goldstone modes in the symmetry broken phase of the corresponding model, i.e. $M=0$ for the chiral Ising model, $M=1$ for the chiral XY model and $M=2$ for the chiral Heisenberg model.
For clarity, however, we list all the beta and gamma functions separately in the following.

\subsection{Chiral Ising model}

For the chiral Ising model, the loop contributions to the beta function of the Yukawa coupling up to three-loop order explicitly read\cite{Mihaila:2017ble}
\begin{align}
	\beta_{y, \chi\text{I}}^{\text{(1L)}}&=(3+2N)y^2\,,\\
	\beta_{y, \chi\text{I}}^{\text{(2L)}}&=24y\lambda(\lambda- y)-\big(\frac{9}{8}+6N\big)y^3\,,\\
	\beta_{y, \chi\text{I}}^{\text{(3L)}}&=\frac{y}{64}\Big(1152 (7+5N) y^2\lambda+192 (91-30 N) y \lambda ^2\nonumber\\
	&+\big(912 \zeta_3-697+2 N (67+112 N+432 \zeta_3)\big)y^3\nonumber\\
	&-13824 \lambda ^3\Big)\,.
\end{align}
Here $\zeta_z$ is the Riemann zeta function. Accordingly, the contributions to the beta function for the quartic scalar coupling are given by
\begin{align}
	\beta_{\lambda, \chi\text{I}}^{\text{(1L)}}&=36 \lambda ^2+4N y \lambda -N y^2\,,\\
	\beta_{\lambda, \chi\text{I}}^{\text{(2L)}}&=4 N y^3+7N y^2 \lambda-72N y \lambda ^2-816 \lambda ^3\,,\\
	\beta_{\lambda, \chi\text{I}}^{\text{(3L)}}&=\frac{1}{32} \Big(6912 (145+96 \zeta_3) \lambda ^4+49536 N y \lambda ^3\nonumber\\
	&-48 N (72 N-361-648 \zeta_3)y^2\lambda ^2\nonumber\\
	&+2 N (1736 N-4395-1872 \zeta_3)y^3 \lambda\nonumber\\
	&+N (5-628 N-384 \zeta_3)y^4\Big)\,.
\end{align}
The four-loop order contributions are quite lengthy and are given in App.~\ref{app:fourloopI}. 
For the contributions to the gamma function corresponding to wave function renormalization of the fermion derivative term, we find
\begin{align}
	\gamma_{\psi, \chi\text{I}}^{\text{(1L)}}&=\frac{y}{2}\,,\\
	\gamma_{\psi, \chi\text{I}}^{\text{(2L)}}&=-\frac{y^2}{16} (12 N+1)\,,
\end{align}
\begin{align}
	\gamma_{\psi, \chi\text{I}}^{\text{(3L)}}&=\frac{y^3}{128}\left(48 \zeta _3+4 N (47-12 N)-15\right)\nonumber\\
	&\quad+6 \lambda  y^2-\frac{33 \lambda ^2 y}{2}\,.
\end{align}
The gamma function corresponding to the wave function renormalization of the derivative term of the scalar order parameter reads
\begin{align}
	\gamma_{\phi, \chi\text{I}}^{\text{(1L)}}&=2Ny\,,\\
	\gamma_{\phi, \chi\text{I}}^{\text{(2L)}}&=24 \lambda ^2-\frac{5 N y^2}{2}\,,\\
	\gamma_{\phi, \chi\text{I}}^{\text{(3L)}}&=-216 \lambda ^3+\frac{1}{32} N y^3 \left(48 \zeta _3+200 N+21\right)\nonumber\\
	&\quad+30 \lambda  N y^2-90 \lambda ^2 N y\,.
\end{align}
Finally, the scaling of the quadratic scalar operator is given by the following contributions to the gamma function $\gamma_{\phi^2, \chi\text{I}}$,
\begin{align}
	\gamma_{\phi^2, \chi\text{I}}^{\text{(1L)}}&=-12\lambda\,,\\
	\gamma_{\phi^2, \chi\text{I}}^{\text{(2L)}}&=144 \lambda ^2-2 N y (y-12 \lambda )\,,\\
	\gamma_{\phi^2, \chi\text{I}}^{\text{(3L)}}&=\frac{3}{2} Ny^2\lambda  \left(24 N-120 \zeta _3-11\right)-6264 \lambda ^3\nonumber\\
	&\quad-4 N y^3 \left(4 N-9+3 \zeta _3\right)-288 N y \lambda ^2\,.
\end{align}
The four-loop order contributions are displayed in App.~\ref{app:fourloopI}. 

This completes the set of beta and gamma functions that are required to determine the fixed-points and the critical exponents for the chiral Ising model.
Our expressions fully agree up to three loops with the ones from our Ref.~\onlinecite{Mihaila:2017ble}. Upon setting $y=0$, the beta function for the quartic coupling also agrees with the four-loop results for the real scalar $\phi^4$ theory with $Z_2$ or Ising symmetry\cite{Kleinert:2001ax}.
As a further check we later compare our four-loop results for the critical exponents
$\eta_\phi$, $\eta_\psi,1/\nu$ and $\omega$ with the large-$N$ results of the GN
model from Refs.~\onlinecite{Gracey:1990wi,Gracey:1992cp,Derkachov:1993uw,Vasiliev:1992wr,Vasiliev:1993pi,Gracey:1993kb,Gracey:2017fzu} and find them to agree.

\subsubsection{Remarks on emergent supersymmetric theory}

For $N=1/4$, the field content of the chiral Ising model is compatible with an
emergent supersymmetry scenario as discussed in
Ref.~\onlinecite{Grover:2013rc}. 
Up to three loops all supersymmetric relations hold exactly~\cite{doi:10.1093/ptep/ptw120,Zerf:2016fti,Mihaila:2017ble}. 
At fourth order the na\"ive $N=1/4$ limit yields a violation of the superscaling relations, because the beta functions for the couplings $y$ and $\lambda$ are not equal upon the rescaling $y\to 8 \lambda$. 
This implies that one of the supersymmetric scaling relations\cite{Gies:2009az,Heilmann:2014iga} between the critical exponents will also be violated at fourth order in $\epsilon$.
However, the original SUSY Lagrangian containing  a two-component Majorana
fermion as the superpartner of a single real scalar was formulated in $D=3$
dimensions\cite{Grover:2013rc,Iliesiu:2015akf}. Up to three loops, one may
perform a $D=4-\epsilon$ dimensional calculation with a four-component Dirac
fermion and formally continue the results to $N=1/4$. At four-loop order, however,
we proved by explicit calculation that the differences in the $\gamma$~algebra in the underlying four-dimensional and three-dimensional cases manifest in the renormalization of the fermion-fermion-scalar vertex. 
This can be explained as follows:

When dimensional regularization is used, the spacetime dimension $D$ becomes
non-integer  and the basis of $\gamma$-matrices needs to be extended to an
infinite-dimensional set\cite{Bondi:1989nq,Gracey:2016mio} 
 \begin{align}\label{eq:prod}
\Gamma^{(k)}_{\mu_1\mu_2\ldots\mu_k} &=\frac{1}{k\,!}
\gamma_{[\mu_1}\gamma_{\mu_2}\cdots \gamma_{\mu_k ]}\quad \mbox{for}\quad k\ge 0\,,
\end{align}
where the square brackets denote the antisymmetrization. Furthermore, we
follow the usual procedure\cite{collins} and impose  the restriction that when
$\epsilon\to 0$ the familiar relations, which are valid in $D=4$ or $D=3$, are restored. 
For example, when the spacetime
dimension $D$ is an integer, the product in Eq.~\eqref{eq:prod} in which each $\gamma$~matrix occurs only once
plays a special role and we denote it  by
 \begin{align}
\Gamma^{(D)}_{\mu_1\mu_2\ldots\mu_D} &\equiv \tilde{\gamma}_D = \gamma_{_1}\gamma_{_2}\cdots
\gamma_{D}\,.
\end{align}
For the matrices $\tilde{\gamma}_D$ it holds
 \begin{align}\label{eq:coeffid}
\tilde{\gamma}_D^2&=\alpha_D^2 I\quad\mbox{with}\nonumber\\
  \alpha_D &=1\quad \mbox{for}\quad D=0,1 \mod 4\,,\nonumber\\
  \alpha_D &=i\quad \mbox{for}\quad D=2,3 \mod 4\,.
\end{align}
Thus, for $D=4$ the more familiar $\gamma_5$ is recovered through the relation
 $\gamma_5 =  \tilde{\gamma}_4$. When continuing to $D=4-\epsilon$ or
$D=3-\epsilon$ we maintain all the properties of the matrices
$\tilde{\gamma}_4$ and $\tilde{\gamma}_3$ from the underlying integer
dimensions.

 The general
strategy is to decompose products of $\gamma$~matrices by the iterative
use  of the following identities
 \begin{align}
\gamma_{\mu}\Gamma^{(k)}_{\mu_1\mu_2\ldots\mu_k} &=
\Gamma^{(k+1)}_{\mu\mu_1\mu_2\ldots\mu_k} \nonumber\\
& +\sum_{i=1}^k (-1)^{i+1}
\delta_{\mu\mu_i}\Gamma^{(k-1)}_{\mu_1\mu_2\ldots\mu_{i-1}\mu_{i+1}\ldots\mu_k}\,,\nonumber\\
\Gamma^{(k)}_{\mu_1\mu_2\ldots\mu_k} \gamma_{\mu}
&=\Gamma^{(k+1)}_{\mu_1\mu_2\ldots\mu_k\mu}\nonumber\\
&+\sum_{i=1}^k (-1)^{n-i}
\delta_{\mu\mu_i}\Gamma^{(k-1)}_{\mu_1\mu_2\ldots\mu_{i-1}\mu_{i+1}\ldots\mu_k}\,.
\end{align}
For $k=1$ one obtains the simple relation
\begin{align}
\gamma_{\mu}\gamma_{\nu}=\Gamma^{(2)}_{\mu\nu}+\delta_{\mu\nu} I =
\frac{1}{2}\left(\gamma_{\mu}\gamma_{\nu}-\gamma_{\nu}\gamma_{\mu}\right)+\delta_{\mu\nu}
I\,, 
\end{align}
where $I$ is the identity matrix. Another useful result is 
\begin{align}
\gamma_{\mu}\Gamma^{(k)}\gamma_{\mu} & = (-1)^k (D-2 k) \Gamma^{(k)}
\label{eq:trd}
\end{align}
from which it follows 
\begin{align}
\operatorname{Tr}(\Gamma^{(k)}) & = 0\quad \mbox{for}\quad \Gamma^{(k)}\ne I,
\tilde{\gamma}_D\,,\nonumber\\
\operatorname{Tr}(\tilde{\gamma}_D) & = 0 \quad \mbox{for} \quad D \quad\mbox{even}\,.
\end{align}
Therefore, when one performs traces over products of $\gamma$~matrices without
$\tilde{\gamma}_D$, one picks out the coefficient of the identity matrix and (if
$D$ is odd) of the matrix $\tilde{\gamma}_D$. As can be understood from
Eq.~\eqref{eq:trd} the trace over a chain of three different $\gamma$~matrices
will be set to zero in $D=4$. In $D=3$ it holds the relation 
\begin{align}
\operatorname{Tr}(\gamma_\mu\gamma_\nu\gamma_\rho) &= \pm i
\varepsilon_{\mu\nu\rho}  \operatorname{Tr}(I)\,,
\end{align}
where $\varepsilon_{\mu\nu\rho}$ denotes the Levi-Civita tensor in three
dimensions. When $D$ is continued to $D=4-\epsilon$ within DREG the properties of
an even dimensional spacetime are preserved and the trace over a chain of
three different $\gamma$~matrices also vanishes. In our setup we slightly
modify DREG and  apply a
semi-na\"ive  regularization prescription\cite{Mihaila:2012pz} for $\tilde{\gamma}_3$ consisting in
the formal replacement
\begin{align}
\operatorname{Tr}(\gamma_\mu\gamma_\nu\gamma_\rho) &= \pm i
\tilde{\varepsilon}_{\mu\nu\rho}  \operatorname{Tr}(I) +{\cal O}(\epsilon)\,.
\label{eq:trndim}
\end{align}
The object $\tilde{\varepsilon}_{\mu\nu\rho}$ has some similarities with the
three-dimensional Levi-Civita tensor: (i) it is
completely antisymmetric in all indices; (ii) when contracted with a second one of its kind
we demand the following result
\begin{align}
\tilde{\varepsilon}^{\mu\nu\rho}
\tilde{\varepsilon}_{\mu^\prime\nu^\prime\rho^\prime}&=\delta^{[\mu}_{[\mu^\prime}
    \delta^{\nu}_{\nu^\prime}\delta^{\rho]}_{\rho^\prime]}\,,
\end{align}
where the square brackets denote complete anti-symmetrization. When  $D= 3$, $\tilde{\varepsilon}^{\mu\nu\rho}$ resembles the
three-dimensional Levi-Civita tensor. To avoid confusion we call this 
prescription $\mbox{DREG}_3$.

\begin{figure}[t]
  \begin{center}
   \includegraphics[width=\columnwidth]{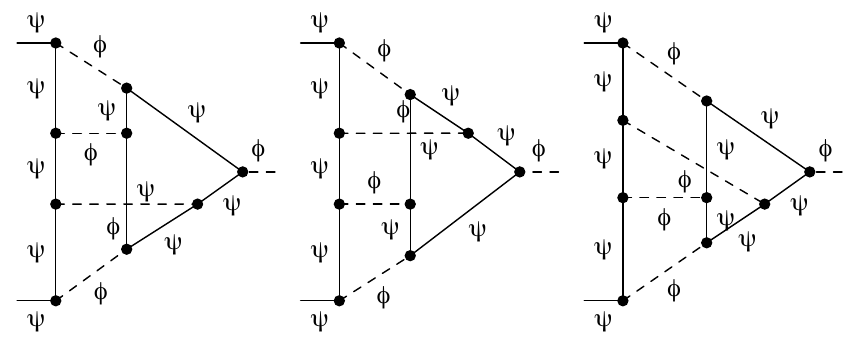}
      \caption[]{\label{fig:diagr}\sloppy Sample diagrams
        contributing to the fermion-fermion-scalar renormalization in $D=3$
        dimensions. The scalar field is denoted $\phi$ and the Majorana
        fermions $\Psi$. }
  \end{center}
\end{figure}

 As the Eq.~\eqref{eq:trndim}
 can only be defined up to an ambiguity of order  ${\cal O}(\epsilon)$, we
 made sure that the four-loop diagrams containing two fermion chains each made
 up of at least three different $\gamma$~matrices  contribute at most simple
 poles in $\epsilon$. Sample diagrams are shown in Fig.~\ref{fig:diagr}.  
After taking into account these contributions the renormalization constant for
the fermion-fermion-scalar vertex gets an additional contribution
proportional to the number of fermions $N$ that we marked with the label $\Delta_3$
in Eq.~\eqref{eq:beta4lci} and that restores the supersymmetric relations.

As a cross check of the method we implemented  the $SU(2)$ spin algebra as
an explicit representation of the Clifford algebra in $D=3$ dimensions for the
four-loop diagrams. With this approach we obtained  complete agreement for the
contributions labeled with $\Delta_3$ in the previous setup.
It is also important to mention that together with SUSY restoration 
the numerical values of the couplings at the fixed point change too and 
 the critical exponents satisfy all the superscaling relations as will be
 discussed in the next section.

\subsection{Chiral XY model}

In this section, we give the RG beta and gamma functions for the chiral XY model. Again, we forward the lengthy four-loop contributions to the appendix, cf. App.~\ref{app:fourloopXY}.
The contributions to the beta function of the Yukawa coupling up to three-loop order explicitly read
\begin{align}
	\beta_{y, \chi\text{XY}}^{\text{(1L)}}&=(2+2N) y^2\,,\\
	\beta_{y, \chi\text{XY}}^{\text{(2L)}}&= 32 \lambda ^2 y+\big(\frac{7}{2}-6 N\big) y^3-32 \lambda\,  y^2\,,\\
	\beta_{y, \chi\text{XY}}^{\text{(3L)}}&=\frac{1}{8} (N (52 N+15)-227) y^4\nonumber\\
	&+24 \lambda  (5 N+6) y^3-120 \lambda ^2 (N-3) y^2\,.\nonumber\\
	&-320 \lambda ^3 y+6 \zeta _3 (N+1) y^4\,.
\end{align}
The contributions to the beta function for the quartic scalar coupling are
\begin{align}
	\beta_{\lambda, \chi\text{XY}}^{\text{(1L)}}&=40 \lambda ^2+4N\lambda y-N y^2\,,\\
	\beta_{\lambda, \chi\text{XY}}^{\text{(2L)}}&=2 N y \left(-40 \lambda ^2+2 y^2+\lambda  y\right)-960 \lambda ^3\,,\\
	\beta_{\lambda, \chi\text{XY}}^{\text{(3L)}}&=39488 \lambda ^4+\frac{1}{8} N (53-154 N) y^4\nonumber\\
	&+1808 \lambda ^3 N y+2 \lambda ^2 N (509-60 N) y^2\nonumber\\
	&-12 \zeta _3 \left(N y^2 \left(-68 \lambda ^2+y^2+7 \lambda  y\right)-2048 \lambda ^4\right)\nonumber\\
	&+\frac{1}{4} \lambda  N (512 N-1339) y^3\,.
\end{align}
For the contributions to the gamma function of the wave function renormalization of the fermion derivative term, we obtain
\begin{align}
	\gamma_{\psi, \chi\text{XY}}^{\text{(1L)}}&=y\,,\\
	\gamma_{\psi, \chi\text{XY}}^{\text{(2L)}}&=-\frac{y^2}{4} (6 N+1) \,,\\
	\gamma_{\psi, \chi\text{XY}}^{\text{(3L)}}&=-\frac{1}{16} (2 N (6 N-37)+35) y^3+3 \zeta _3 y^3\nonumber\\
	&+16 \lambda  y^2-44 \lambda ^2 y\,,
\end{align}
The gamma function which corresponds to the wave function renormalization of the derivative term of the complex order parameter reads
\begin{align}
	\gamma_{\phi, \chi\text{XY}}^{\text{(1L)}}&=2N y\,,\\
	\gamma_{\phi, \chi\text{XY}}^{\text{(2L)}}&=32\lambda^2-3Ny^2\,,\\
	\gamma_{\phi, \chi\text{XY}}^{\text{(3L)}}&=-320 \lambda ^3+6 \zeta _3 N y^3+\frac{1}{8} N (64 N-27) y^3\nonumber\\
	&+40 \lambda  N y^2-120 \lambda ^2 N y\,,
\end{align}
The scaling of the wave function renormalization of the squared mass term of the complex order parameter field is given by the contributions
\begin{align}
	\gamma_{\phi^2, \chi\text{XY}}^{\text{(1L)}}&=-16\lambda\,,\\
	\gamma_{\phi^2, \chi\text{XY}}^{\text{(2L)}}&=32 \lambda  (6 \lambda +N y)\,,\\
	\gamma_{\phi^2, \chi\text{XY}}^{\text{(3L)}}&=-4 \Big(2336 \lambda ^3+N (6 N-11) y^3+96 \lambda ^2 N y\nonumber\\
	&+6 \zeta _3 N y^2 (8 \lambda +y)+\lambda  N (25-12 N) y^2\Big)\,.
\end{align}
The four-loop order contributions are displayed in App.~\ref{app:fourloopXY}. This completes the set of beta and gamma functions that are required to determine the fixed-points and the critical exponents for the chiral XY model.

We compare these results on the two-loop level, with the ones for the
NJL-Yukawa model as given in Ref.~\onlinecite{doi:10.1093/ptep/ptw120} and
confirm, that they are in complete agreement. Further, in the case $N=1/2$, we
check that at the three-loop level all the beta and gamma functions coincide
with the ones given previously in Ref.~\onlinecite{Zerf:2016fti}. For $y=0$,
we recover the corresponding four-loop expressions for the bosonic $\phi^4$
theory with $O(2)$ symmetry\cite{Kleinert:2001ax}. Further, consistency checks
based on exact results for the critical exponents and SUSY relations for the
case $N=1/2$ can be found in the results section. Let us mention at this
point, that there is no additional contribution when going from
$D=4-\epsilon$ to $D=3-\epsilon$ spacetime dimenensions, because diagrams
similar with those shown in Fig.~\ref{fig:diagr} do not exist in the chiral XY
  model. Therefore, the  SUSY relations can be easily obtained from our results via the formal
  limit $N=1/2$.

\subsection{Chiral Heisenberg model}

For the chiral Heisenberg model, the contributions to the beta function of the Yukawa coupling explicitly read
\begin{align}
	\beta_{y, \chi\text{H}}^{\text{(1L)}}&=(1+2N) y^2\,,\\
	\beta_{y, \chi\text{H}}^{\text{(2L)}}&=y \left(40 \lambda ^2+\left(\frac{47}{8}-6 N\right) y^2-40 \lambda  y\right)\,,
\end{align}
\begin{align}
	\beta_{y, \chi\text{H}}^{\text{(3L)}}&=\frac{1}{64} \big(608 N^2+278 N-2731\big) y^4\nonumber\\
	&+\frac{9}{4} \zeta _3 (7-2 N) y^4+150 \lambda  (N+1) y^3\,\nonumber\\
	&+5 \lambda ^2 (89-30 N) y^2-440 \lambda ^3 y
\end{align}
Accordingly, the contributions to the beta function for the quartic bosonic coupling are
\begin{align}
	\beta_{\lambda, \chi\text{H}}^{\text{(1L)}}&=44 \lambda ^2+4Ny\lambda-N y^2\,,\\
	\beta_{\lambda, \chi\text{H}}^{\text{(2L)}}&=N y \left(-88 \lambda ^2+4 y^2-3 \lambda  y\right)-1104 \lambda ^3\,,\\
	\beta_{\lambda, \chi\text{H}}^{\text{(3L)}}&=48184 \lambda ^4+\frac{1}{32} N (365-604 N) y^4\nonumber\\
	&+\frac{1}{16} \lambda  N (2360 N-6339) y^3\nonumber\\
	&+3 \zeta _3 \left(9472 \lambda ^4+N y^2 \left(212 \lambda ^2-4 y^2-19 \lambda  y\right)\right)\nonumber\\
	&+\frac{1}{2} \lambda ^2 N (3067-264 N) y^2+2068 \lambda ^3 N y\,.
\end{align}
For the gamma function corresponding to wave function renormalization of the fermion derivative term, we find
\begin{align}
	\gamma_{\psi, \chi\text{H}}^{\text{(1L)}}&=\frac{3y}{2}\,,\\
	\gamma_{\psi, \chi\text{H}}^{\text{(2L)}}&=-\frac{9}{16} (4 N+1) y^2\,,\\
	\gamma_{\psi, \chi\text{H}}^{\text{(3L)}}&=\frac{3}{128} y \big(\left(-48 N^2+404 N-183\right) y^2\nonumber\\
	&-3520 \lambda ^2+240 \zeta _3 y^2+1280 \lambda  y\big)\,.
\end{align}
The gamma function contributions to the wave function renormalization of the boson derivative term read
\begin{align}
	\gamma_{\phi, \chi\text{H}}^{\text{(1L)}}&=2Ny\,,\\
	\gamma_{\phi, \chi\text{H}}^{\text{(2L)}}&=40 \lambda ^2-\frac{7 N y^2}{2}\,,\\
	\gamma_{\phi, \chi\text{H}}^{\text{(3L)}}&=-440 \lambda ^3+\frac{15}{2} \zeta _3 N y^3+\frac{1}{32} N (312 N-131) y^3\nonumber\\
	&+50 \lambda  N y^2-150 \lambda ^2 N y\,.
\end{align}
Finally, the scaling of the wave function renormalization of the squared mass term of the bosonic order parameter field is given by the following contributions
\begin{align}
	\gamma_{\phi^2, \chi\text{H}}^{\text{(1L)}}&=-20\lambda\,,\\
	\gamma_{\phi^2, \chi\text{H}}^{\text{(2L)}}&=240 \lambda ^2+2 N y (20 \lambda +y)\,,\\
	\gamma_{\phi^2, \chi\text{H}}^{\text{(3L)}}&=-12920 \lambda ^3+8 N (7-4 N) y^3-480 \lambda ^2 N y\nonumber\\
	&-36 \zeta _3 N y^2 (5 \lambda +y)+\frac{5}{2} \lambda  N (24 N-89) y^2\,.
\end{align}
All four-loop order contributions are given in App.~\ref{app:fourloopH}.

We benchmark our results on a two-loop level with the ones given in Ref.~\onlinecite{Rosenstein:1993zf}. We have detected a mismatch, which is hard to track down, as in Ref.~\onlinecite{Rosenstein:1993zf} only the final results are listed for the chiral Heisenberg model.
Eventually, the mismatch only shows up in the $\epsilon^2$ coefficient of the inverse correlation length exponent, cf. Sec.~\ref{ref:resH}. 
All the other exponents agree up to two-loop order.
After careful cross-checking we are confident that the results presented here are correct.

\section{Critical exponents}\label{sec:results}

The four-loop beta functions allow the determination of the RG fixed points of the system order by order in $\epsilon$ up to $\mathcal{O}(\epsilon^4)$.
At the one-loop level, the beta functions for $y$ and $\lambda$ give rise to four different fixed points: the unstable Gau\ss ian fixed point with vanishing coordinates $(y_\ast,\lambda_\ast)_0=(0,0)$, the unstable bosonic Wilson-Fisher fixed point $(y_\ast,\lambda_\ast)_\text{WF}=\left(0,\epsilon/(4M+36)\right)$, and a pair of fully non-Gau\ss ian fixed points~(NGFP)
\begin{align}\label{eq:NGFP}
(y_\ast,\lambda_\ast)_{\pm}=\left(\frac{\epsilon}{3+2N-M},\frac{\frac{1}{8}(3-2N-M\pm s)\epsilon}{(3+2 N-M)(M+9)}\right)\,,\nonumber
\end{align}
where
\begin{align}
s=\sqrt{9+4N(33+N)+20 N M+M^2-6 M}\,.
\end{align}
From the pair of NGFPs, the one with the negative solution has a negative quartic coupling. In the following, we do not discuss this fixed-point. 
We note, however, that it has been considered in the context of conformal field theories\cite{doi:10.1093/ptep/ptw120,Iliesiu:2017nrv}.
Here, we study the stable positive solution from Eq.~\eqref{eq:NGFP}, only, which we solve order by order in $\epsilon$.

The universal critical exponents which we determine, here, are the (inverse) correlation length  exponent $\nu^{-1}$, the anomalous dimensions of bosons and fermions, $\eta_\phi$ and $\eta_\psi$, respectively, as well as the subleading exponent $\omega$.
To obtain the fermion and boson anomalous dimensions, we evaluate the gamma functions $\gamma_\psi$ and $\gamma_\phi$ at the corresponding NGFP, i.e.,
\begin{align}
\eta_\psi&=\gamma_\psi(y_\ast,\lambda_\ast)\,,\\ 
\eta_\phi&=\gamma_\phi(y_\ast,\lambda_\ast)\,.
\end{align}
The RG beta function of the dimensionless mass term $\tilde m^2=\mu^{-2}m^2$ follows from Eq.~\eqref{eq:Z} and reads 
\begin{align}
\beta_{\tilde m^2}=(-2+\gamma_\phi-\gamma_{\phi^2})\,\tilde m^2\,.
\end{align}
This beta function is used to extract the inverse correlation length exponent with the relation
\begin{align}
\nu^{-1}= \theta_1=-\frac{d\beta_{\tilde m^2}}{d \tilde m^2}\Big|_{(y^\ast,\lambda^\ast)}=2-\eta_\phi+\eta_{\phi^2}\,,
\end{align}
where, in agreement with the previous notation, we have defined $\eta_{\phi^2}=\gamma_{\phi^2}(y_\ast,\lambda_\ast)$.

Eventually, we access the subleading exponent $\omega$ as the smaller eigenvalue of the stability matrix $\mathcal{M}$, i.e. the matrix of first derivatives of the beta functions with respect to the couplings, evaluated at the stable fixed point
\begin{align}\label{eq:stabmat}
	\mathcal{M}=
	\begin{pmatrix}
	\frac{\partial\beta_{y}}{\partial y} & \frac{\partial\beta_{\lambda}}{\partial y}\\
	\frac{\partial\beta_{y}}{\partial \lambda} & \frac{\partial\beta_{\lambda}}{\partial \lambda}
	\end{pmatrix}\bigg|_{y_\ast,\lambda_\ast}\,.
\end{align}
Then, the smaller eigenvalue $\omega$ corresponds to the less irrelevant RG direction and the larger eigenvalue, which we call $\omega^\prime$, is more irrelevant.
For the corrections to scaling, the less irrelevant contribution is more important and therefore we will only list $\omega$ from now on.

Next, we evaluate the beta and gamma functions for different explicit choices of fermion flavor number $N$ to provide explicit results for the inverse correlation length exponent, subleading exponent and the fermion and boson anomalous dimensions. The full expressions with arbitrary $N$ can be given analytically, however, they are very lengthy and we therefore do not display them, here. 
Instead, we have prepared supplemental material with the full expressions of the critical exponents for general $N$ in three separate files for the three separate models\cite{suppmat}.

\subsection{Chiral Ising model}

Here, we discuss the most important cases of the chiral Ising model, i.e. the semimetal-CDW transition  in graphene ($N=2$), the semimetal-insulator transition of spinless fermions on the honeycomb lattice $(N=1)$ and the emergent SUSY scenario $(N=1/4)$. Finally, we also comment on the limit $N\to 0$.

\begin{table}[t!]
\caption{\label{tab:critexp} {\it Chiral Ising universality} in $D=3$: Inverse correlation length exponent $1/\nu$ and anomalous dimensions  $\eta_\phi$  and $\eta_\psi$ for bosons and fermions, respectively. In {\it this work}, we provide results within the $(4-\epsilon)$ expansion to $\mathcal{O}(\epsilon^4)$. We do not give values for critical exponents where the Pad\'e approximant contains a pole in the interval $D \in [2,4]$. To obtain the values for $1/\nu$ from the QMC results, we have performed simple numerical inversions of the values for $\nu$ given in the corresponding references.}
\begin{tabular*}{\linewidth}{@{\extracolsep{\fill} } c c c c c c}
\hline\hline
$N=1/4$  & & $1/\nu$ & $\eta_\phi$ & $\eta_\psi$ & $\omega$ \\
\hline
{\it this work}, $P_{[2/2]}$ & & 1.415 & 0.171 & 0.171 & 0.843\\
{\it this work}, $P_{[3/1]}$ & & 1.415 & 0.170 & 0.170 & 0.838\\[8pt]
FRG\cite{Gies:2017tod} (Regulator 1) & &  1.385 & 0.174 & 0.174 & 0.765\\
FRG\cite{Gies:2017tod} (Regulator 2) & &  1.395 & 0.167 & 0.167 & 0.782\\
conformal bootstrap\cite{Iliesiu:2015akf}  & &  & 0.164 & 0.164 & \\
\hline
$N=1$  & & $1/\nu$ & $\eta_\phi$ & $\eta_\psi$ & $\omega$\\
\hline
{\it this work}, $P_{[2/2]}$ & & -  & 0.4969 & 0.0976 & 0.779 \\
{\it this work}, $P_{[3/1]}$ & & -  & 0.4872 & 0.0972 & 0.760 \\
{\it this work}, $P_{[0/4]}$ & & 1.101 & - & - & - \\[8pt]
FRG\cite{PhysRevB.94.245102} & & 1.075(4) & 0.5506 & 0.0645 &  \\
conformal bootstrap\cite{Iliesiu:2017nrv}  & & 0.76 & 0.544 & 0.084 &  \\
Monte Carlo\cite{1367-2630-17-8-085003} & & 1.30 & 0.45(3) &  &  \\
Monte Carlo\cite{Huffman:2017swn} & & 1.14 & 0.54(6) &  &  \\
\hline
$N=2$  & & $1/\nu$ & $\eta_\phi$ & $\eta_\psi$ & $\omega$ \\
\hline
{\it this work}, $P_{[2/2]}$ & & 0.931 & 0.7079 & 0.0539 & 0.794 \\
{\it this work}, $P_{[3/1]}$ & & 0.945 & 0.6906 & 0.0506 & 0.777 \\[8pt]
$(2+\epsilon)$, ($\epsilon^4$, Pad\'e)\cite{Gracey:2016mio}& & 0.931 & 0.745 & 0.082 & \\
FRG\cite{PhysRevB.94.245102}  & & 0.994(2) & 0.7765 & 0.0276 & \\
conformal bootstrap\cite{Iliesiu:2017nrv}  & & 0.88 & 0.742 &  0.044 & \\
Monte Carlo\cite{PhysRevD.88.021701} & & 1.20(1) & 0.62(1) & 0.38(1) & \\
\hline\hline
\end{tabular*}
\end{table}
%

For $N=2$, the numerical evaluation of the critical exponents provides the following series in $\epsilon$,
\begin{align}
	\frac{1}{\nu}&\approx 2-0.9524 \epsilon+0.007225 \epsilon ^2-0.09487 \epsilon ^3-0.01265 \epsilon ^4\,,\nonumber\\
	\eta_\phi&\approx 0.5714 \epsilon+0.1236 \epsilon ^2-0.02789 \epsilon ^3+0.1491 \epsilon ^4\,\nonumber\,,\\
	\eta_\psi&\approx 0.07143 \epsilon-0.006708 \epsilon ^2-0.02434 \epsilon ^3+0.01758 \epsilon ^4\,,\nonumber\\
	\omega&\approx \epsilon-0.3525 \epsilon ^2+0.4857 \epsilon ^3-1.338 \epsilon ^4\,.
\end{align}
The full analytical expression for this series is given in App.~\ref{app:fourloopI}. We note that the second order coefficient in the series for the inverse correlation length exponent and in the one for the fermion anomalous dimension seem to be accidentally small. 

To obtain first estimates for the critical exponents for the physical case of (2+1) dimensions, we employ simple Pad\'e approximants and note that a more thorough analysis of resummations and interpolations is underway. 
The results from the Pad\'e estimates are listed in Tab.~\ref{tab:critexp}, together with the estimates from other approaches, i.e. the $2+\epsilon$ expansion\cite{GRACEY1990403,Gracey:2016mio}, the functional RG\cite{PhysRevB.94.245102}, the conformal bootstrap\cite{Iliesiu:2017nrv} and quantum Monte Carlo\cite{PhysRevD.88.021701}.
We have chosen to display the symmetric Pad\'e approximant $P_{[2/2]}$ as well as $P_{[3/1]}$ for comparison.
The results from the other available Pad\'e approximants, i.e.  $P_{[4/0]},P_{[1/3]},P_{[0/4]}$, are distributed in a larger interval, tentatively contain poles in $D \in\{2,4\}$ for some of the critical exponents or are ill-defined.
This observation proliferates to the other values of $N$ studied here as well as to the chiral XY and chiral Heisenberg model.
Therefore, we do not display them in the following, except for special cases.
We postpone a more thorough study of this matter to future work.
In summary, we can see that for $N=2$, the estimates of the field-theoretical approaches agree rather well for the inverse correlation length exponent and the boson anomalous dimension, in particular when focusing on the two different $\epsilon$ expansions. On the other hand, the uncertainty in the determination of the fermion anomalous dimension remains rather large. The quantum Monte Carlo results for $1/\nu$ and $\eta_\psi$ are also quite far from the other approaches and it will be an interesting task to track the origin for that difference in the future.

In Tab.~\ref{tab:critexp}, we have also listed the results for the intensely studied $N=1$ case.
The Pad\'e approximants for critical exponents which contain a pole in the interval $D \in\{2,4\}$ should be interpreted carefully and we therefore refrain from displaying them in the tables.
We find that for the inverse correlation length exponent all the Pad\'e approximants have poles in $D \in\{2,4\}$, except for $P_{[0/4]}$, which we have additionally listed in Tab.~\ref{tab:critexp}.
Comparisons to the other approaches are also listed in Tab.~\ref{tab:critexp}, exhibiting good agreement for the boson anomalous dimension. The results for $1/\nu$, however, are scattered over a rather large interval.
Here, the estimates from the two different RG approaches\cite{PhysRevB.94.245102} are located approximately half way between the estimates from quantum Monte Carlo\cite{1367-2630-17-8-085003} and the conformal bootstrap\cite{Iliesiu:2017nrv}.

As already pointed out, for $N=1/4$ the field content of the chiral Ising model is compatible with an
emergent supersymmetric model. After taking into account the contributions
occurring solely in the $D=3$ computation, we find for the  critical
exponents the following series
\begin{align}
 \frac{1}{\nu}=& 2-\frac{4 \epsilon }{7}-\frac{\epsilon ^2}{49}+\left(\frac{60 \zeta _3}{2401}-\frac{29}{4802}\right) \epsilon ^3\\
 &+\frac{\left(-40 \zeta _3-5000 \zeta _5+14 \pi ^4-141\right) \epsilon ^4}{67228}+\mathcal{O}(\epsilon^5)\,,\nonumber\\
 \eta=& \frac{\epsilon }{7}+\frac{2 \epsilon ^2}{49}+\left(\frac{29}{2401}-\frac{120 \zeta _3}{2401}\right) \epsilon ^3\nonumber\\
 &+\frac{\left(40 \zeta _3+5000 \zeta _5-14 \pi ^4+141\right) \epsilon ^4}{33614}+\mathcal{O}(\epsilon^5)\,,
 \end{align}
with $\eta=\eta_\psi=\eta_\phi$ and
 \begin{align}
 &\omega = \epsilon-\frac{3 \epsilon ^2}{7}+\left(\frac{282 \zeta _3}{343}+\frac{129}{686}\right) \epsilon ^3\\
 &+\frac{\left(329 \pi ^4-42660 \zeta_3-158400 \zeta_5-9075\right) \epsilon ^4}{48020}+\mathcal{O}(\epsilon^5)\,.\nonumber
\end{align}
Thus, the superscaling relation\cite{Gies:2009az,Heilmann:2014iga} 
\begin{align}
\frac{1}{\nu}&=\frac{D-\eta}{2}\,,
\end{align}
exactly holds up to fourth order in $\epsilon$. Numerically we obtain
\begin{align}
\frac{1}{\nu}=\frac{D-\eta}{2}\approx\ & 2. -0.571429 \epsilon-0.0204082 \epsilon^2\nonumber\\
&+0.0239998 \epsilon^3-0.0596477 \epsilon^4\,.
\end{align}
Pad\'e approximants for the critical exponents in 2+1 dimensions are given in Tab.~\ref{tab:critexp}. We observe a good agreement between the available estimates from the conformal bootstrap approach\cite{Iliesiu:2015akf}, the FRG\cite{Gies:2017tod} and our results.

Finally, we would like to comment on the replica limit for our GNY model, $N\to 0$, which has been argued to be applicable to the transition from a relativistic semi-metallic state to a diffusive metallic phase in a 3D Weyl semi-metal\cite{PhysRevB.94.220201}.
A four-loop expansion of the purely fermionic Gross-Neveu model in $D=2+\epsilon$ exhibits large $\mathcal{O}(\epsilon^4)$ contributions when compared to the three-loop contributions\cite{Gracey:2016mio}.
This is attributed to the fact that the limit $N\to 0$ suffers from the lack of multiplicative renormalizability yielding contributions from evanescent operators\cite{PhysRevB.94.220201,Gracey:2016mio}.
It was argued in Ref.~\onlinecite{PhysRevB.94.220201} that this problem can be circumvented by considering the GNY model in the same limit.
At one-loop order the $N\to 0$ limit of the GNY model gives rise to a NGFP which is non-trivial in both couplings $(y^\ast,\lambda^\ast)_+=(\epsilon/3,\epsilon/36)$.
This is exactly the fixed point which has been considered in Ref.~\onlinecite{PhysRevB.94.220201}.
We remark, however, that for $N\to 0$ the beta function $\beta_{\lambda, \chi\text{I}}$ and the gamma functions $\gamma_{\phi, \chi\text{I}}, \gamma_{\phi^2, \chi\text{I}}$ completely decouple from the fermionic sector order by order in the loop expansion as every term proportional to the squared Yukawa coupling $y$ comes at least with a factor~$N$.
Therefore, the critical exponents $\nu^{-1}, \eta_\phi$ and $\omega$ are identical to the ones obtained for the purely bosonic Ising field theory\cite{Kleinert:2001ax} which we confirm by evaluating our equations in this limit to order $\mathcal{O}(\epsilon^4)$. These exponents are amended by a non-trivial fermion anomalous dimension, which we display here for the sake of completeness,
\begin{align}
	\frac{1}{\nu}&= 2-\frac{1}{3}\epsilon -\frac{19}{162} \epsilon ^2+\left(\frac{4 \zeta _3}{27}-\frac{937}{17496}\right) \epsilon ^3\nonumber\\
	&+\frac{\left(771120 \zeta _3-4665600 \zeta _5+11664 \pi ^4-124285\right) }{9447840}\epsilon ^4\nonumber\\
	&+\mathcal{O}(\epsilon^5)\,,
\end{align}
and
\begin{align}
	\eta_\phi&= \frac{1}{54}\epsilon ^2+\frac{109}{5832}\epsilon ^3+\left(\frac{7217}{629856}-\frac{4 \zeta _3}{243}\right) \epsilon ^4+\mathcal{O}(\epsilon^5)\,,\\
	\eta_\psi&= \frac{1}{6}\epsilon +\frac{31}{648} \epsilon ^2+\left(\frac{1475}{34992}-\frac{2 \zeta _3}{27}\right) \epsilon ^3\nonumber\\
	&+\frac{\left(1944000 \zeta _5-338580 \zeta _3-5832 \pi ^4-8465\right)}{9447840}\epsilon^4\nonumber\\
	&+\mathcal{O}(\epsilon^5)\,.
\end{align}
Due to the effective relativistic invariance of this model the dynamical critical exponent is $z=1$. Together with the conventional Ising critical exponent which is known with great accuracy, this is in stark contrast to numerical data\cite{PhysRevLett.112.016402,sbierski2015}. We therefore conjecture that the universality class of the semimetallic state to a diffusive metallic phase in a 3D Weyl semimetal is likely to be different than the one from the replica limit of the chiral Ising model.

\subsection{Chiral XY model}

In the chiral XY model there are two specific values for the number of fermion flavors $N$ which are relevant to condensed-matter applications, i.e. the quantum transition of surface states in topological insulators as covered by the choice $N=1/2$ and the superconducting transition in graphene where $N=2$. Further, the case $N=2$ is relevant to a Kekul\'e transition in graphene which is described by a complex $Z_3$ order parameter, however, exhibits emergent $U(1)$ symmetry at the QCP.

Here, we start with the discussion of $N=1/2$ which has been conjectured to exhibit an emergent supersymmetry at the QCP.
We find the critical exponents
\begin{align}
	\frac{1}{\nu}&= 2-\epsilon+\frac{\epsilon ^2}{3}-\left(\frac{2 \zeta _3}{3}+\frac{1}{18}\right) \epsilon ^3\nonumber\\
	&+\frac{1}{540}\left(420 \zeta_3+1200 \zeta_5-3 \pi ^4+35\right) \epsilon ^4+\mathcal{O}(\epsilon^5)\,,\\[8pt]
	\eta_\phi &=\eta_\psi =\epsilon/3+\mathcal{O}(\epsilon^5)\,,
\end{align}
and for the subleading exponent
\begin{align}
	\omega&=\epsilon-\frac{\epsilon ^2}{3}+\left(\frac{2 \zeta_3}{3}+\frac{1}{18}\right) \epsilon ^3\nonumber\\
	&-\frac{1}{540}\left(420 \zeta_3+1200 \zeta_5-3 \pi ^4+35\right) \epsilon ^4+\mathcal{O}(\epsilon^5)\,.
\end{align}
For this case, there is a supersymmetric scaling relation\cite{thomas2005}, connecting the correlation length exponent and the subleading exponent, reading 
\begin{align}
	\nu^{-1}=2-\omega\,.
\end{align}
Comparing the above equations, we confirm that this relation is exactly
fulfilled order by order in the $\epsilon$ expansion through four loops.

Further, as in the case of the chiral Ising model, we provide Pad\'e approximants to obtain estimates for the critical exponents in $D=2+1$. These are listed in Tab.~\ref{tab:critexpXY} together with the result from the conformal bootstrap approach\cite{Bobev:2015vsa} exhibiting good agreement between the different methods for $N=1/2$. 
We also note that the result for the anomalous dimensions $\eta_\phi=\eta_\psi=1/3$ agrees exactly with the one-loop result up to $\mathcal{O}(\epsilon^4)$. This is in agreement with SUSY non-renormalization theorems\cite{strassler2003}.

\begin{table}[t!]
\caption{\label{tab:critexpXY} {\it Chiral XY universality} in $D=3$: Inverse correlation length exponent $1/\nu$ and anomalous dimensions  $\eta_\phi$  and $\eta_\psi$ for bosons and fermions, respectively. In {\it this work}, we provide results within the $(4-\epsilon)$ expansion to order~$\mathcal{O}(\epsilon^4)$.}
\begin{tabular*}{\linewidth}{@{\extracolsep{\fill} } c c c c c c}
\hline\hline
$N=1/2$  & & $1/\nu$ & $\eta_\phi$ & $\eta_\psi$ & $\omega$ \\
\hline
{\it this work}, $P_{[2/2]}$ & & 1.128 & 1/3 & 1/3 & 0.872\\
{\it this work}, $P_{[3/1]}$ & & 1.130 & 1/3 & 1/3 & 0.870\\[8pt]
conformal bootstrap\cite{Bobev:2015vsa}  & & 1.090 & 1/3 & 1/3 & 0.910\\
\hline
$N=2$  & & $1/\nu$ & $\eta_\phi$ & $\eta_\psi$ & $\omega$\\
\hline
{\it this work}, $P_{[2/2]}$ & & 0.840 & 0.810 & 0.117 & 0.796 \\
{\it this work}, $P_{[3/1]}$ & & 0.841 & 0.788 & 0.108 & 0.780 \\[8pt]
functional RG\cite{Classen:2017hwp} & & 0.862 & 0.88 & 0.062 & 0.878 \\
Monte Carlo\cite{li2015} & & 1.06(5) & 0.71(3) &  &  \\
\hline\hline
\end{tabular*}
\end{table}
%

For $N=2$ the numerical evaluation of the critical exponents gives
\begin{align}
	\frac{1}{\nu}&\approx 2-1.2 \epsilon+0.1829 \epsilon^2-0.3515 \epsilon^3+0.5164 \epsilon^4\,,\nonumber\\
	\eta_\phi&\approx 0.6667 \epsilon+0.1211 \epsilon ^2-0.005048 \epsilon ^3+0.1938 \epsilon ^4\,\nonumber\,,\\
	\eta_\psi&\approx 0.1667 \epsilon-0.02722 \epsilon ^2-0.05507 \epsilon ^3+0.04202 \epsilon ^4\,,\nonumber\\
	\omega&\approx \epsilon-0.3783 \epsilon ^2+0.6271 \epsilon ^3-1.853 \epsilon ^4\,.
\end{align}
The corresponding Pad\'e approximants are shown in Tab.~\ref{tab:critexpXY} and the full analytical expressions can be found in App.~\ref{app:fourloopXY}. 
As before, we only give the results for the Pad\'e approximants $P_{[2/2]}$ and $P_{[3/1]}$ which, in the case of the chiral XY model, do not show any poles for $D\in \{2,4\}$ for investigated values of $N$.
In Tab.~\ref{tab:critexpXY}, we also provide the estimates from the functional RG\cite{Classen:2017hwp} and recent quantum Monte Carlo calculations\cite{li2015}.
Again, the results for the inverse correlation length exponent agree reasonably well within the different RG approaches, however, there is a rather large difference when compared to the QMC results.
It will be interesting to see estimates from the conformal bootstrap approach for this case.

\subsection{Chiral Heisenberg model}\label{ref:resH}

Finally, we discuss the chiral Heisenberg model for eight-component spinors, i.e. $N=2$, which corresponds to the field-theoretical formulation of the antiferromagnetic transition of interacting electrons on the honeycomb lattice as relevant to graphene and related materials. 
For the inverse correlation length exponent, the boson and fermion anomalous dimension and the subleading exponent, we find the numerical results
\begin{align}
	\frac{1}{\nu}&\approx 2-1.527 \epsilon+0.4076 \epsilon ^2-0.8144 \epsilon ^3+2.001 \epsilon ^4 \,,\nonumber\\
	\eta_\phi&\approx 0.8 \epsilon+0.1593 \epsilon ^2+0.02381 \epsilon ^3+0.2103 \epsilon ^4\,\nonumber\,,\\
	\eta_\psi&\approx 0.3 \epsilon-0.05760\epsilon^2-0.1184 \epsilon^3+0.04388 \epsilon^4\,,\nonumber\\
	\omega&\approx \epsilon-0.4830 \epsilon^2+0.9863 \epsilon^3-2.627 \epsilon^4\,.
\end{align}
The full analytical expressions are given in App.~\ref{app:fourloopH}.
We note that the second order coefficient of the inverse correlation length
exponent is different from the one given by Rosenstein,
cf.~Refs.~\onlinecite{PhysRevB.89.205403,Rosenstein:1993zf}. 
After careful checks, doing two independent calculations and exploiting two
different ways for determining the renormalization constant for the mass term,
we come to the conclusion that our results are correct.
Let us also notice that a mistake in the two-loop calculation usually shows up
as non-local or divergent  contributions to the beta-functions or the
anomalous dimensions at three and four loops. Our results do not contain
such problematic contributions, that reassures the consistency of our
results. Furthermore, we verified that when changing the underlying $SU(2)$ symmetry to
 $U(1)$ symmetry we recover the results for the chiral Ising model.  The expressions for the boson
and fermion anomalous dimensions agree with the two-loop results from
Ref.~\onlinecite{Rosenstein:1993zf}. 

\begin{table}[t!]
\caption{\label{tab:critexpH} {\it Chiral Heisenberg universality} in $D=3$: Inverse correlation length exponent $1/\nu$ and anomalous dimensions  $\eta_\phi$  and $\eta_\psi$ for bosons and fermions, respectively. In {\it this work}, we provide results within the $(4-\epsilon)$ expansion to order~$\epsilon^4$. We do not give values for critical exponents where the Pad\'e approximant contains a pole in the interval $D \in [2,4]$. The values for $1/\nu$ printed in italic are simple numerical inversions of the values for $\nu$ as given in the corresponding references.}
\begin{tabular*}{\linewidth}{@{\extracolsep{\fill} } c c c c c c}
\hline\hline
$N=2$  & & $1/\nu$ & $\eta_\phi$ & $\eta_\psi$ & $\nu$ \\
\hline
{\it this work} (Pad\'e [2/2]) & & 0.6426 & 0.9985 & 0.1833 & -\\
{\it this work} (Pad\'e [3/1]) & & 0.6447 & 0.9563 & 0.1560 & 1.2352\\[8pt]
functional RG\cite{Knorr:2017yze} & &  0.795 & 1.032 & 0.071 & 1.26\\
Monte Carlo\cite{Otsuka:2015iba} & &  {\it 0.98} &  & 0.20(2) & 1.02(1)\\
Monte Carlo\cite{toldin2014} & &  {\it 1.19} & 0.70(15) & & 0.84(4)\\
\hline\hline
\end{tabular*}
\end{table}
%

In Tab.~\ref{tab:critexpH}, we provide estimates for the critical exponents in $D=2+1$ dimensions from Pad\'e approximants. Also, we list the values found by other approaches, i.e. a recent functional RG calculation\cite{Knorr:2017yze} and the quantum Monte Carlo approach\cite{Otsuka:2015iba} to the semimetal-insulator transition of interacting lattice electrons with massless Dirac-like dispersion relations.
We observe that the different approaches do not show a satisfactory agreement for the critical exponents. The deviation of the estimates for the inverse correlation length exponent between the pRG and FRG approaches are of order 20\% and the distance to the QMC result is even bigger.

Alternatively, the inverted series of $\nu^{-1}$ can also be considered in order to obtain a more direct estimate for the correlation length exponent. 
The series reads
\begin{align}
	\nu\approx 0.5 +0.3818 \epsilon+0.1897 \epsilon^2+0.2706 \epsilon^3-0.1768\epsilon^4.\nonumber
\end{align}
The Pad\'e approximants for the series in $\nu$ evaluated at $\epsilon=1$ are also given in Tab.~\ref{tab:critexpH}.
We note that this improves the comparison with the FRG approach which also agrees well on the boson anomalous dimension. On the other hand it does not resolve the rather large difference to the numerical estimate from the QMC simulations.

\section{Conclusions} 

We have studied the chiral Ising, the chiral XY and the chiral Heisenberg model at four-loop order in $D=4-\epsilon$ space-time dimensions and have extracted the solutions of the stable non-Gau\ss ian fixed point as well as the corresponding critical exponents to order~$\mathcal{O}(\epsilon^4)$.
Further, we have calculated simple Pad\'e approximants to provide estimates for the critical exponents in $2+1$ dimensions.
The models investigated are relevant to quantum transitions in a number of condensed-matter physics applications recently discussed in the context of Dirac and Weyl semimetals.
For the first time, we give the full analytical expressions for the beta and gamma functions for the chiral Ising, XY and Heisenberg models for general number of fermion flavors $N$ at the four-loop level. 
Explicitly, we calculated the inverse correlation length exponent, the subleading exponent and the anomalous dimensions for bosons as well as fermions for specific quantum phase transitions in different condensed-matter and field-theoretical setups. The relevant applications include interaction-induced transitions in graphene and other Dirac materials, surface states of topological insulators and the emergence supersymmetric quantum critical conformal field theories.

For the chiral Ising model at $N=1/4$, we observe good agreement of the estimates for critical exponents across different field-theoretical methods, including perturbative RG, functional RG and the conformal bootstrap. 
For this scenario, emergent supersymmetry at the quantum critical point has been conjectured and, here, we have confirmed that the supersymmetric scaling relations hold up to four-loop order in the perturbative RG approach.
For $N=2$, as relevant for interacting electrons in graphene, the agreement between the different renormalization group methods turns out to be reasonable, however, the results from the QMC simulations deviate significantly. 
Also, there is a large difference as compared to recent conformal bootstrap results which deviate from the QMC simulations even more strongly.
This issue remains to be resolved.
The chiral XY model at $N=1/2$ exactly fulfills the corresponding supersymmetric scaling relation order by order and the critical exponents are found to be in good agreement with the conformal bootstrap results.
For the chiral XY and Heisenberg models at $N=2$, we observe again a reasonable agreement between the different RG approaches at least for some critical exponents, but a rather unsatisfactory gap as compared to the lattice results.

It will be interesting to track down the origin of the remaining deviations between the different approaches in the future. One possible origin of the differences might be effects from corrections to scaling\cite{toldin2014,ayyar2015} at least when it comes to the comparison between the renormalization group and the lattice methods. For example, a close to marginal scaling of the difference between the boson and fermion velocities which is generally non-vanishing in the lattice approaches could be difficult to assess and have a strong impact on the fitting procedure of the appropriate scaling functions.
Within the perturbative RG approach a thorough analysis of resummation and interpolation techniques is certainly required which we postpone to future work.

\paragraph*{Acknowledgments}
The authors are grateful to John Gracey, Bernhard Ihrig, Lukas Janssen, Achim
Rosch and Bj\"orn Sbierski for discussions and John Gracey for reading the manuscript. L.M. would like to thank
Konstantin Chetyrkin for enlightening conversations about the technical aspects of the calculation. 

\appendix

\begin{widetext}
\section{Four-loop contributions for the chiral Ising model}\label{app:fourloopI}

The four-loop contributions to the beta functions read
\begin{align}\label{eq:beta4lci}
	\beta_{y, \chi\text{I}}^{\text{(4L)}}&=-\frac{5}{2} \zeta _5 (42 N+43) y^5+\frac{\left(32 \pi ^4 (2 N+3) (18 N+19)+40 N (8 N (44 N-899)+29721)+457935\right) y^5}{7680}\\
	&+\frac{\lambda}{8} (8 N (12 N-683)-2829) y^4-\frac{1}{2} \lambda ^2 (4 N (6 N+635)+4455) y^3+36 \lambda ^3 (8 N-455) y^2\nonumber\\
	&-\frac{1}{8} \zeta _3 y^2 \left(-41472 \lambda ^3+(4 N (125 N+331)-5)
        y^3+432 \lambda  (12 N+7) y^2-864 \lambda ^2 (6 N-25) y\right)+14040
        \lambda ^4 y\,,\nonumber\\
& + \Delta_3 N (1 + 107 \zeta _3 - 125  \zeta _5) y^5\,,\nonumber
\end{align}
\begin{align}
	\beta_{\lambda, \chi\text{I}}^{\text{(4L)}}=&41472 \left(-39 \zeta _3-60 \zeta _5+\frac{\pi ^4}{10}-\frac{3499}{96}\right) \lambda ^5+\frac{1}{240} \lambda  N y^4 \left(-60 \zeta _3 \left(912 N^2-4156 N-4677\right)\right.\\
	&\left.+1200 \zeta _5 (157-168 N)-4 \pi ^4 (450 N+41)+25 (4 N (337 N+3461)+5847)\right)\nonumber\\
	&+\frac{N y^5 \left(480 \zeta _3 (12 N (14 N-15)+277)+2400 \zeta _5 (128 N+65)+8 \pi ^4 (64 N-77)+160 N (1289-386 N)-67095\right)}{1920}\nonumber\\
	&+\frac{1}{80} \lambda ^2 N y^3 \left(835200 \zeta _5+1920 \zeta _3 (3 N (4 N-61)+19)+72 \pi ^4 (24 N+31)-40 N (288 N+15649)+1057825\right)\nonumber\\
	&+\frac{4}{5} \lambda ^3 N y^2 \left(-86400 \zeta _5+540 \zeta _3 (4
        N-69)+7890 N-288 \pi ^4-72605\right)+\frac{36}{5} \left(-17280
        \zeta_3+96 \pi ^4-6775\right) \lambda ^4 N y\,.\nonumber
\end{align}
The symbol $\Delta_3$ should be set to $\Delta_3=1$ in $\mbox{DREG}_3$,
{\it e.g.} together with $N=1/4$  the limit of an emergent
supersymmetric theory is recovered.  For the generic case of DREG in 
$D=4-2\epsilon$ dimensions    it holds $\Delta_3=0$.

The four-loop contributions to the gamma functions read
\begin{align}
	\gamma_{\psi, \chi\text{I}}^{\text{(4L)}}=&\frac{y}{393216} \Bigg(134479872 \lambda ^3+y^3 \left(-884736 \zeta _3-5 \left(384 \left(256 \zeta _5-893\right)+\frac{377339 \pi ^4}{90}\right)\right.\\
	&\left.+16 N \left(-164352 \zeta _3+N \left(1536 \left(16 \zeta _3-3\right) N-74752\right)-\frac{1536 \pi ^4}{5}+53440\right)+\frac{303611 \pi ^4}{18}\right)\nonumber\\
	&-288 \lambda  y^2 \left(512 \left(93-32 \zeta _3\right)+8 \left(7424+\frac{927 \pi ^4}{4}\right) N-\frac{1}{18} \pi ^4 (33372 N+7079)+\frac{7079 \pi ^4}{18}\right)\nonumber\\
	&+96 \lambda ^2 y \left(221184 \zeta _3+344064 N-656384\right)\Bigg)\,,\nonumber
\end{align}
\begin{align}
	\gamma_{\phi, \chi\text{I}}^{\text{(4L)}}&=14040 \lambda ^4+\frac{1}{256} \lambda  N y^3 \left(768 \left(16 \zeta _3-83\right)-19456 N\right)+\frac{1}{32} \lambda ^2 N y^2 \left(256 \left(81 \zeta _3-91\right)-384 N\right)+288 \lambda ^3 N y\\
	&-\frac{N y^4 \left(377856 \zeta _3+15360 \left(8 \zeta _5-29\right)+4 N \left(162816 \zeta _3+256 \left(144 \zeta _3-101\right) N+\frac{1536 \pi ^4}{5}-54016\right)+1024 \pi ^4\right)}{24576}\,,\nonumber
\end{align}
\begin{align}
	\gamma_{\phi^2, \chi\text{I}}^{\text{(4L)}}&=1728 \left(18 \zeta _3+\frac{2 \pi ^4}{5}+187\right) \lambda ^4+\frac{3}{2} \lambda ^2 N y^2 \left(5760 \zeta _3+4 \left(-48 \zeta _3-176\right) N+\frac{48 \pi ^4}{5}+3796\right)\\
	&+\frac{1}{64} N y^4 \left(-5376 \zeta _3+10080 \zeta _5+2 N \left(320 \zeta _3+4480 \zeta _5+64 \left(18 \zeta _3-11\right) N+\frac{48 \pi ^4}{5}-5208\right)-\frac{224 \pi ^4}{5}-2846\right)\nonumber\\
	&-\frac{3}{16} \lambda  N y^3 \left(-5120 \zeta _3-5760 \zeta _5+4 N \left(-672 \zeta _3+64 \left(2 \zeta _3-1\right) N+\frac{16 \pi ^4}{3}-1618\right)+\frac{184 \pi ^4}{5}+12989\right)\nonumber\\
	&+36 \left(96 \zeta _3+313\right) \lambda ^3 N y\,.\nonumber
\end{align}
%

\subsection{Critical exponents for the chiral Ising model for $N=2$}

The full analytical expressions for the most important critical exponents for the chiral Ising model at $N=2$ read
\begin{align}
	\frac{1}{\nu} =& 2-\frac{20\epsilon}{21}+\frac{325\epsilon^2}{44982}-\frac{(271572144\zeta_3+36133009)\epsilon ^3}{3821940612}\nonumber\\
	&+\frac{\left(73192843310400 \zeta _3+179520471709200 \zeta _5-2472257012904 \pi ^4-86141171013035\right)\epsilon ^4}{4175164363361040}+\mathcal{O}(\epsilon^5)\,,
\end{align}
\begin{align}
	\eta_\phi =& \frac{4 \epsilon }{7}+\frac{109 \epsilon ^2}{882}+\left(\frac{1170245}{26449416}-\frac{144 \zeta _3}{2401}\right) \epsilon ^3\nonumber\\
	&+\frac{\left(162669869280 \zeta _3+171915696000 \zeta _5-1203409872 \pi ^4+102456536695\right) \epsilon ^4}{2407822585560}+\mathcal{O}(\epsilon^5)\,,\\
	\eta_\psi =& \frac{\epsilon }{14}-\frac{71 \epsilon ^2}{10584}-\left(\frac{18 \zeta _3}{2401}+\frac{2432695}{158696496}\right) \epsilon ^3\nonumber\\
	&+\frac{\left(1155813964920 \zeta _3+515747088000 \zeta _5-3610229616 \pi ^4-556332486445\right) \epsilon ^4}{57787742053440}+\mathcal{O}(\epsilon^5)\,,\\
	\omega=& \epsilon-\frac{533 \epsilon ^2}{1512}+\left(\frac{165 \zeta _3}{686}+\frac{6685099}{34006392}\right) \epsilon ^3\nonumber\\
	&+\frac{\left(-46250341862688 \zeta _3-23579956863360 \zeta (5)+119137577328 \pi ^4-11065294400875\right) \epsilon ^4}{59438820397824}+\mathcal{O}(\epsilon^5)\,.
\end{align}
%

\section{Four-loop contributions for the chiral XY model}\label{app:fourloopXY}

The four-loop contributions to the beta functions of the chiral XY model read
\begin{align}
	\beta_{y, \chi\text{XY}}^{\text{(4L)}}&=-\frac{8}{3} \lambda ^2 \left(6 N^2+508 N+1257\right) y^3-40 \zeta _5 (N+1) y^5+\frac{2}{3} \lambda  (8 N (3 N-208)+99) y^4\nonumber\\
	&+\frac{1}{480} \left(32 \pi ^4 (N+1)^2-20 N (4 N (3 N+619)-6703)+20295\right) y^5+\frac{320}{3} \lambda ^3 (4 N-215) y^2\nonumber\\
	&-\zeta _3 y^2 \left(-6144 \lambda ^3+(6 N (5 N+28)-163) y^3+32 \lambda  (18 N+11) y^2+64 \lambda ^2 (53-9 N) y\right)+21120 \lambda ^4 y\,,\\[2pt]
	\beta_{\lambda, \chi\text{XY}}^{\text{(4L)}}&=2 \zeta _3 \Big(-1052672 \lambda ^5+N \left(22 N^2-56 N+21\right) y^5+\lambda  N (2 N (285-62 N)+557) y^4\nonumber\\
	&+32 \lambda ^2 N (N (5 N-63)+25) y^3+64 \lambda ^3 N (16 N-175) y^2-72704 \lambda ^4 N y\Big)\nonumber\\
	&+10 \zeta _5 \left(-311296 \lambda ^5+N (16 N+5) y^5+8 \lambda  N (7-12 N) y^4+1248 \lambda ^2 N y^3-8192 \lambda ^3 N y^2\right)\nonumber\\
	&+\frac{1}{240} \Big(20480 \left(64 \pi ^4-23925\right) \lambda ^5+N \left(8 \pi ^4 (4 N-7)-5 (4 N (436 N-1977)+5229)\right) y^5\nonumber\\
	&+8 \lambda  N \left(5 N \left(915 N-36 \pi ^4+7910\right)+36 \pi ^4+16935\right) y^4\nonumber\\
	&+8 \lambda ^2 N \left(-600 N (8 N+489)+8 \pi ^4 (68 N+69)+514455\right) y^3\nonumber\\
	&+512 \lambda ^3 N \left(3505 N-128 \pi ^4-43845\right) y^2+512 \left(384 \pi ^4-28195\right) \lambda ^4 N y\Big)\,.
\end{align}
The four-loop contributions to the gamma functions read
\begin{align}
	\gamma_{\psi, \chi\text{XY}}^{\text{(4L)}}&=\frac{1}{2} \zeta _3 y^2 \left(192 \lambda ^2+\left(4 N^3-49 N-17\right) y^2+64 \lambda  y\right)+\frac{1}{960} \left(32 \pi ^4 (N+1)-20 N (6 N (3 N+76)-47)+20055\right) y^4\nonumber\\
	&-\frac{2}{3} \lambda  (174 N+143) y^3+\frac{4}{3} \lambda ^2 (168 N-277) y^2-20 \zeta _5 y^4+\frac{3040 \lambda ^3 y}{3}\,,\\
	\gamma_{\phi, \chi\text{XY}}^{\text{(4L)}}&=21120 \lambda ^4+\zeta _3 N y^2 \left(576 \lambda ^2-\left(4 N^2+30 N+31\right) y^2+64 \lambda  y\right)-40 \zeta _5 N y^4+\frac{1280}{3} \lambda ^3 N y\nonumber\\
	&+\frac{1}{60} N \left(15 (N-38) N+4 \pi ^4 (N+1)+3120\right) y^4-\frac{8}{3} \lambda  N (38 N+137) y^3-\frac{16}{3} \lambda ^2 N (3 N+130) y^2\,,\\
	\gamma_{\phi^2, \chi\text{XY}}^{\text{(4L)}}&=80 \zeta _5 N y^3 (16 \lambda +3 N y+2 y)+4 \zeta _3 \Big(12800 \lambda ^4+N (2 N (6 N-5)-17) y^4-32 \lambda  (N-6) N (N+2) y^3\nonumber\\
	&+32 \lambda ^2 N (77-3 N) y^2+1280 \lambda ^3 N y\Big)+\frac{1}{30} \Big(4096 \left(3765+8 \pi ^4\right) \lambda ^4+N \left(4 \pi ^4 (N-7)-20 N (48 N+197)+405\right) y^4\nonumber\\
	&+2 \lambda  N \left(-8 \pi ^4 (8 N+11)+120 N (8 N+225)-56915\right) y^3+64 \lambda ^2 N \left(-660 N+7 \pi ^4+5900\right) y^2+503680 \lambda ^3 N y\Big)\,.\nonumber
\end{align}
%

\subsection{Critical exponents for the chiral XY model for $N=2$}

The full analytical expressions for the most important critical exponents for $N=2$ in the chiral XY model read
\begin{align}
	\frac{1}{\nu} &= 2-\frac{6 \epsilon }{5}+\frac{823 \epsilon
          ^2}{4500}+\left(\frac{909821}{12150000}-\frac{5986          \zeta_3}{16875}\right) \epsilon ^3\nonumber\\ 
	&\quad+\frac{\left(245618820 \zeta_3+1924948800 \zeta_5-9697320 \pi ^4+347495879\right) \epsilon ^4}{3280500000}+\mathcal{O}(\epsilon^5)\,,\\[8pt]
\eta_\phi&=\frac{2 \epsilon }{3}+\frac{109 \epsilon ^2}{900}-\frac{1363 \epsilon ^3}{270000}+\left(\frac{300977 \zeta _3}{2025000}+\frac{1103491}{72900000}\right) \epsilon ^4+\mathcal{O}(\epsilon^5)\,,\\[8pt]
\eta_\psi&=\frac{\epsilon }{6}-\frac{49 \epsilon ^2}{1800}-\frac{29737 \epsilon ^3}{540000}+\left(\frac{226913 \zeta _3}{4050000}-\frac{1477451}{58320000}\right) \epsilon ^4+\mathcal{O}(\epsilon^5)\,,\\[8pt]
\omega&=\epsilon-\frac{227 \epsilon ^2}{600}+\left(\frac{\zeta _3}{6}+\frac{4801}{11250}\right) \epsilon ^3-\frac{\left(1043012880 \zeta _3+270000000 \zeta (5)-1350000 \pi ^4+398475259\right) \epsilon ^4}{972000000}+\mathcal{O}(\epsilon^5)\,.\nonumber
\end{align}

\section{Four-loop contributions to chiral Heisenberg}\label{app:fourloopH}

The four-loop contributions to the beta functions read
\begin{align}
	\beta_{y, \chi\text{H}}^{\text{(4L)}}&=-\frac{5}{2} \lambda ^2 \left(8 N^2+508 N+1949\right) y^3+\frac{5}{2} \zeta _5 (29-62 N) y^5+\frac{5}{8} \lambda  (8 N (4 N-327)+1241) y^4+\frac{20}{3} \lambda ^3 (88 N-4475) y^2\nonumber\\
	&-\frac{\left(96 \pi ^4 (2 N-7) (2 N+1)+40 N (8 N (68 N+3971)-104923)+1096225\right) y^5}{7680}+29000 \lambda ^4 y\nonumber\\
	&+\frac{1}{8} \zeta _3 y^2 \left(53760 \lambda ^3+(4 N (7 N-269)+1437) y^3-80 \lambda  (36 N+43) y^2+160 \lambda ^2 (18 N-193) y\right)\,,\\[8pt]
	\beta_{\lambda, \chi\text{H}}^{\text{(4L)}}&=\frac{5}{4} \zeta _5 \left(-3022848 \lambda ^5+N (128 N+23) y^5+36 \lambda  N (11-24 N) y^4+9824 \lambda ^2 N y^3-75776 \lambda ^3 N y^2\right)\nonumber\\
	&+\frac{1}{4} \zeta _3 \Big(-10631168 \lambda ^5+N (4 N (46 N-185)+251) y^5+\lambda  N (4 N (1283-268 N)+4775) y^4\nonumber\\
	&+32 \lambda ^2 N (N (44 N-449)+478) y^3+64 \lambda ^3 N (148 N-779) y^2-661504 \lambda ^4 N y\Big)\nonumber\\
	&+\frac{1}{1920}\Big(2048 \left(6512 \pi ^4-2473805\right) \lambda ^5-N \left(320 N (243 N-1409)+88 \pi ^4+400085\right) y^5\nonumber\\
	&+8 \lambda  N \left(100 N (395 N+2661)+4 \pi ^4 (137-266 N)+125685\right) y^4\nonumber\\
	&+8 \lambda ^2 N \left(-120 N (352 N+23653)+8 \pi ^4 (424 N+315)+5249585\right) y^3\nonumber\\
	&+512 \lambda ^3 N \left(32410 N-1184 \pi ^4-504075\right) y^2+512 \left(3552 \pi ^4-269585\right) \lambda ^4 N y\Big)\,.
\end{align}
The four-loop contributions to the gamma functions read
\begin{align}
	\gamma_{\psi, \chi\text{H}}^{\text{(4L)}}&=\frac{3}{16} \zeta _3 y^2 \left(480 \lambda ^2+\left(16 N^3-235 N-56\right) y^2+320 \lambda  y\right)-\frac{5}{8} \lambda  (348 N+293) y^3+\frac{5}{4} \lambda ^2 (336 N-467) y^2\nonumber\\
	&+\frac{\left(96 \pi ^4 (6 N-1)-40 N (8 N (9 N+310)+1525)+273125\right) y^4}{5120}-\frac{165 \zeta _5 y^4}{4}+2090 \lambda ^3 y\,,\\[8pt]
	\gamma_{\phi, \chi\text{H}}^{\text{(4L)}}&=29000 \lambda ^4-55 \zeta _5 N y^4+\frac{1}{120} N \left(3 \pi ^4 (6 N-1)-5 N (89 N+773)+10885\right) y^4-\frac{5}{3} \lambda  N (76 N+299) y^3\nonumber\\
	&+\frac{1}{8} \zeta _3 N y^2 \left(2880 \lambda ^2-(4 N (4 N+65)+265) y^2+640 \lambda  y\right)-20 \lambda ^2 N (N+26) y^2+\frac{1760}{3} \lambda ^3 N y\,,\\
	\gamma_{\phi^2, \chi\text{H}}^{\text{(4L)}}&=\frac{64}{3} \left(34975+74 \pi ^4\right) \lambda ^4-\frac{1}{480} N \left(16 \pi ^4 (3 N+29)+40 N (504 N+737)-43345\right) y^4\nonumber\\
	&+\frac{5}{2} \zeta _5 N y^3 (720 \lambda +136 N y+13 y)+\frac{1}{48} \lambda  N \left(-8 \pi ^4 (24 N+25)+120 N (32 N+991)-263595\right) y^3\nonumber\\
	&+\frac{10}{3} \lambda ^2 N \left(-528 N+4 \pi ^4+6593\right) y^2+2 \zeta _3 \Big(40000 \lambda ^4+N (5 N (6 N-11)-21) y^4+20 \lambda  N (N (11-4 N)+53) y^3\nonumber\\
	&+80 \lambda ^2 N (64-3 N) y^2+3520 \lambda ^3 N y\Big)+\frac{69580}{3} \lambda ^3 N y\,.
\end{align}
%

\subsection{Critical exponents for the chiral Heisenberg model for $N=2$}

For $N=2$, we find the following critical exponents for the chiral Heisenberg model:
\begin{align}
	\frac{1}{\nu} &= 2-\frac{84 \epsilon }{55}+\frac{2576729 \epsilon ^2}{6322250}+\left(\frac{3834385959243}{13808110112500}-\frac{157961052 \zeta_3}{173861875}\right) \epsilon ^3\nonumber\\
&+\frac{\left(4419355262033682960 \zeta_3+16115053820113182000 \zeta_5-91331632816626840 \pi ^4+11013561507164036543\right) \epsilon ^4}{12063041156482250000}\nonumber\\
&+\mathcal{O}(\epsilon^5)\,,
\end{align}
\begin{align}
\eta_\phi&=\frac{4 \epsilon }{5}+\frac{4819 \epsilon ^2}{30250}+\left(\frac{48 \zeta_3}{625}-\frac{476430591}{6954475000}\right) \epsilon ^3\nonumber\\
&+\frac{\left(310739849149400 \zeta_3+194418190384000 \zeta_5+972090951920 \pi ^4-350384810029259\right) \epsilon ^4}{1518892112375000}+\mathcal{O}(\epsilon^5)\,,\\
\eta_\psi&=\frac{3 \epsilon }{10}-\frac{6969 \epsilon ^2}{121000}+\left(\frac{18 \zeta_3}{625}-\frac{2128383117}{13908950000}\right) \epsilon ^3\nonumber\\
&+\frac{3 \left(406214104344840 \zeta_3+194418190384000 \zeta_5+972090951920 \pi ^4-606847185834529\right) \epsilon ^4}{12151136899000000}+\mathcal{O}(\epsilon^5)\,,\\
\omega&=\epsilon-\frac{11689 \epsilon ^2}{24200}+\left(\frac{27 \zeta _3}{250}+\frac{1191302179}{1390895000}\right) \epsilon ^3\nonumber\\
&+\frac{3 \left(-28239146782436128 \zeta_3-25857619321072000 \zeta_5+20413909990320 \pi ^4-822859847457915\right) \epsilon ^4}{68046366634400000}+\mathcal{O}(\epsilon^5)\,.
\end{align}

\end{widetext}

%

\end{document}